44# The GAPS Time-of-Flight Detector

**Sydney Feldman**[a,*] **for the GAPS collaboration**

[a]*Dept. of Physics and Astronomy, University of California*
*Los Angeles, CA, 90095 USA*

E-mail: sydneyfeldman@physics.ucla.edu
The General Antiparticle Spectrometer (GAPS) Antarctic long duration balloon mission is scheduled for launch during the austral summer of 2024-25. Its novel detection technique, based on exotic atom formation, excitation, and decay, is specifically designed for the detection of slow moving cosmic antiprotons and antideuterons. Such antinuclei are predicted by a wide variety of allowed dark matter models, as well as other astrophysical theories like primordial black holes.

There are two main components of the GAPS instrument: a large-area tracker and a surrounding time-of-flight system (TOF). The combination of these two systems allows GAPS to effectively differentiate between species of negatively-charged antinuclei and determine the energy deposition, velocity, and trajectory of particles interacting with the detector. This contribution will focus on the TOF, which determines the velocity of the incoming antiparticle and provides the trigger to the experiment. An overview of the TOF detector, an explanation of relevant electronics, and a report on its construction and preliminary performance will be discussed. The TOF is composed of 160 thin plastic scintillator paddles ranging in length from 1.5 to 1.8 meters. At each paddle end, signals from six silicon photomultipliers are combined to measure the signal waveforms. These signals are also used to create the trigger. This design is optimized for low mass and fast data acquisition while still maintaining good light collection.
38th International Cosmic Ray Conference (ICRC2023)
26 July - 3 August, 2023
Nagoya, Japan
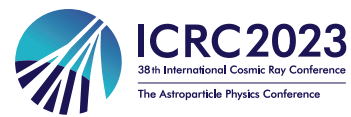

*Speaker

© Copyright owned by the author(s) under the terms of the Creative Commons
Attribution-NonCommercial-NoDerivatives 4.0 International License (CC BY-NC-ND 4.0).                    https://pos.sissa.it/

## 1. Introduction and Science Background

The unknown nature of dark matter (DM) requires investigation with many different experimental strategies. Typically, DM experiments are one of four complementary types: accelerator searches, direct detection, indirect detection, or astrophysical probes. Each of these approaches are optimized for examining the problem of DM from different angles [1]. The General AntiParticle Spectrometer (GAPS) is an indirect detection experiment which targets cosmic ray (CR) antinuclei as a potential dark matter signature. While antiparticles are not themselves DM candidates, a wide range of existing dark matter models predict antinuclei as a product of DM annihilation or decay [2, 3].

For this reason, CR antinuclei hold important information about the question of dark matter. A few years ago, the interpretation of Alpha Magnetic Spectrometer (AMS-02) antiproton data resulted in a potential DM signal (e.g. [4]). However, further analysis [5, 6] reduced the significance of the finding. The difficulty in fully understanding and disentangling background antiprotons from a potential signal motivates the GAPS search for antideuterons.

In addition to extending the antiproton spectrum to lower energies, the GAPS experiment will search this energy range for a potential antideuteron flux. This is a particularly interesting search due to the uniquely low background from known astrophysical processes. According to conventional physics, cosmic ray antideuterons are produced only through interactions of primary CRs with the interstellar medium (ISM). The minimum energy of an incident particle required for this interaction to produce antideuterons is high, and due to relativistic kinematics the resulting spectra of antideuterons (secondary CRs) are boosted to higher energies, with a very low flux predicted at kinetic energies below 0.5 GeV/$n$ [3]. Theories of antideuteron production beyond the standard model are not constrained in the same way, because DM annihilation happens nearly at rest. For example, annihilation of WIMPs would produce an antideuteron flux significantly above conventional physics backgrounds at low energies.

GAPS is a long duration balloon experiment scheduled to launch from Antarctica during the austral summer of 2024-25. This contribution focuses on the time-of-flight system (TOF), which measures energy depositions, particle velocity, and the trajectory of incoming CR particles and provides the trigger to the experiment.

## 2. The GAPS Instrument

The GAPS instrument is composed of two sub detectors: a large area lithium-drifted silicon tracker (Si(Li)), and the time-of-flight detector (TOF) which surrounds it. Detection and particle identification are accomplished with the following novel technique:

When a negatively charged antiparticle enters the detector volume, it will first interact with and deposit energy in the TOF, which forms the trigger for the experiment. Then, it slows down and is captured by the tracker material, forming a matter-antimatter bound state. This state is unstable, and the decay and annihilation of the exotic atom produces X-rays, pions, and protons which also deposit energy in the tracker and TOF. This technique is optimized for the detection of CR antiparticles with less than 0.25 GeV per nucleon, where antiprotons have never been measured and antideuteron backgrounds are extremely low. This technique is less limited in aperture and instrument weight compared to magnetic spectrometers. With over 7 m$^2$ of active Si(Li) detector area and 25 m$^2$ of TOF detector area, the geometric acceptance of GAPS is high, and yet it is light enough to be launched with a long-duration balloon. In addition, the energy of the X-rays are characteristic of the reduced mass of the exotic atom, and can, therefore, be used to determine the charge-to-mass ratio of the captured particle. Crucial identification power comes from the reconstruction of annihilation products [7].

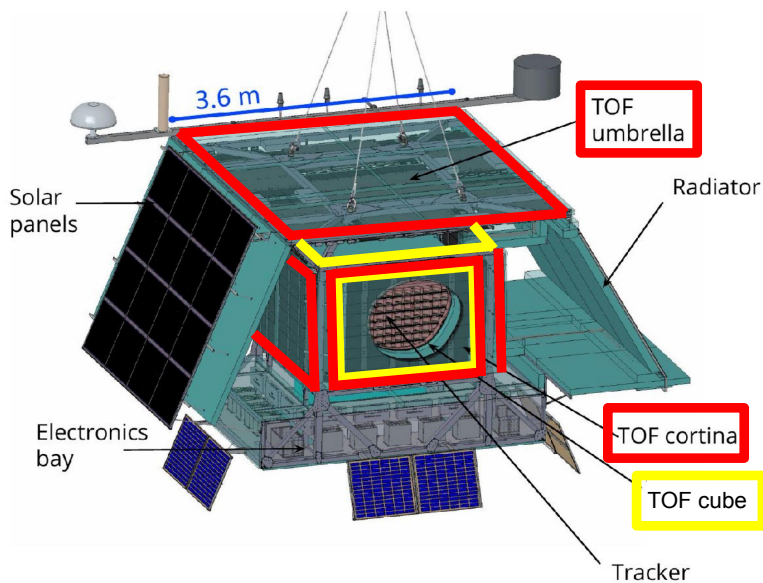

**Figure 1:** Overview of the GAPS instrument, with the outer TOF highlighted in red and the inner TOF highlighted in yellow.

### 2.1 Tracker

The tracker is composed of more than 1000 Si(Li) detectors, each of which is 10 cm in diameter, 2.5 mm thick, and segmented into eight strips. Four of the detectors are grouped together to form a module, which uses an Application Specific Integrated Circuit (ASIC) with 32 channels to provide power and read out data [8]. The modules are arranged in 7 layers with six rows of six modules.

The tracker acts as both the target material for the capture of negatively charged antiparticles, and as an X-ray detector with a resolution of 4 keV (FWHM) in the relevant energy range [9]. The operational range of the tracker is around –35° Celsius, which is accomplished with a passive oscillating heat pipe thermal system [10].

### 2.2 Time-of-Flight Detector (TOF)

The time-of-flight detector (TOF) is composed of 160 long, thin plastic scintillator paddles ranging in length from 1.5 to 1.8 meters. The paddles are arranged with a slight overlap between adjacent paddles into an inner TOF (the cube, closely surrounding the tracker on all six sides) and the outer TOF (a second layer on top of the cube, called the umbrella, and around the four sides, called the cortina). See Figure 1 for a diagram of the experiment with the inner and outer TOFs labeled, and Figure 2 for photos of a TOF paddle and panel.

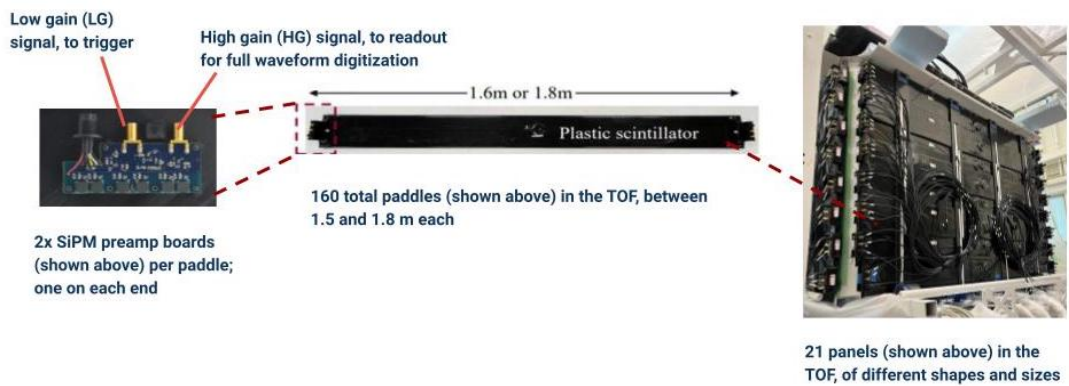

**Figure 2:** A SiPM preamp, fully wrapped paddle, and mounted TOF panel.

This arrangement is designed to provide maximum coverage for events stopping in the tracker. The TOF is also used to track the out-going annihilation products, which is crucial for particle identification.

## 3. TOF Construction

The TOF detector is made out of Eljen EJ-200 general use plastic scintillator. The scintillator was machined by Eljen to have a width of 16 cm, a thickness of 6.35 mm, and lengths varying from 1.5 to 1.8 m. The paddles were wrapped first in aluminum foil and then vinyl blackout material in order to achieve good internal reflection and light-tightness. A small opening was left at the center of the paddle, where a venting tube was inserted in order to allow for outgassing and pressure changes without damage to the wrapping.

The scintillator light output is collected and read out by Silicon photomultipliers (SiPMs). Six SiPMs are attached to each side of a paddle, which connect to a preamplifier board. Aluminum enclosures were built around each SiPM preamp board, and each board was tested to confirm that all six of the SiPMs had a similar gain. For this test, an LED, step motor, and dark box was used to collect data from 1000 LED flashes in front of each of the six SiPMs. Although the results did not provide an exact gain value, they clearly indicated when one of the SiPMs was non-functional, and they also showed the spread in the response of performance across SiPMs (which was determined to be a measure of the relative gain). Examples of acceptable and unacceptable SiPM responses, depicted with distributions of integrated charge per event, are shown in Figure 3.

When a preamp was shown to have an even response to light pulses across all six SiPMs, it was coupled to the end of a paddle using silicon optical interfaces and optical cement. Two preamp boards were adhered to each paddle. An intricate wrapping procedure was then performed to make the nontrivial geometry of the preamp enclosure light-tight. To confirm light-tightness, the paddle was thoroughly inspected using a flashlight, with a specific focus on the edges of the wrapping material and any difficult corners.

Then, muon calibration data were taken for each paddle to verify that the gains were balanced between paddle ends; this ensures uniform illumination of the paddle. Finally, the paddles were strapped to carbon fiber panels which mounted to the gondola frame of the experiment. The panels are arranged into the inner TOF, which is a *cube* that closely encloses the tracker from all six sides, and the outer TOF, which is composed of the *umbrella* (a single layer above the cube top) and the *cortina* (an additional layer around the four cube sides); see Figure 1.

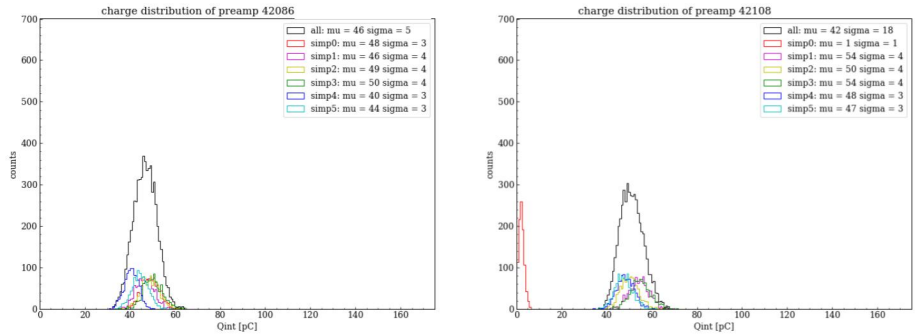

**Figure 3:** Left: Distribution of integrated charge per event for 6000 events with a well-performing SiPM preamp. Right: Distribution of integrated charge per event for an unacceptable SiPM preamp. Note that one SiPM of the six in the preamp did not have a response to the LED pulse.

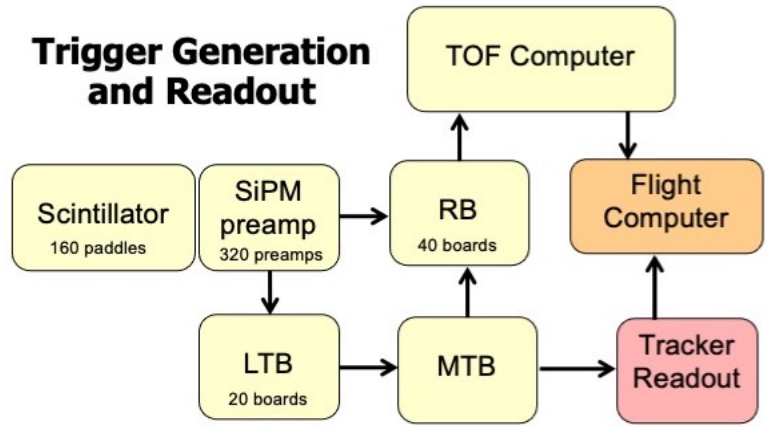

**Figure 4:** A diagram of the signal pathway for the GAPS trigger generation and readout.

Using the time difference between signals from the same paddle along with the known speed of light in the scintillator material, we are able to determine where along the paddle the particle traversed. This knowledge of the hit location, as well as the information available in the signals themselves (like the total integrated charge and pulse height), will give us information on the trajectory, the velocity ($\beta$), and the energy deposited by incoming and outgoing particles. The required timing resolution for the TOF detector is 400ps.

## 4. TOF Electronics

Figure 4 shows the signal pathway for trigger generation and data readout. Each SiPM preamp (two for each of the 160 paddles, for a total of 320 preamps) produces two copies of the signal: a low gain (LG) version and a high gain (HG) version. The LG signal is sent to a local trigger board (LTB), where it is passed through three discriminator levels. The three trigger levels are adjustable, and are currently set to 0.4 min-I (HIT), 2.5 min-I (BETA), and 30 min-I (VETO). HIT is intended to record all paddles with a track in them. BETA is intended to record paddles which have a slow-moving, high-ionizing track. VETO is intended to reject high-Z cosmic ray nuclei, such as Carbon and above. Each LTB encodes the signal from eight paddles using these three trigger levels, and sends a serialized data stream to the master trigger board (MTB) indicating the thresholds reached. There are 20 total LTBs, and the MTB takes the hit pattern from all 20 to make a trigger decision.

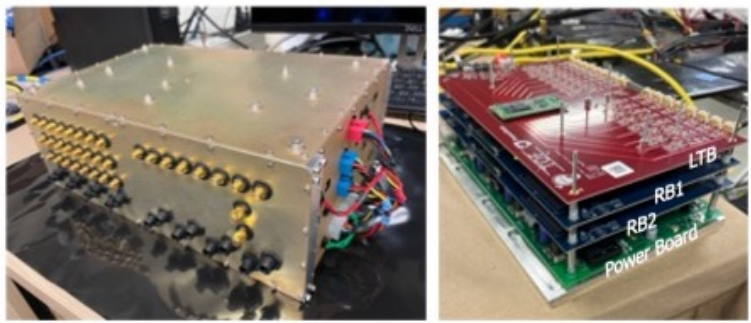

**Figure 5:** Readout and trigger box, closed (left) and partially disassembled (right) to show the electronics boards within.

The tentative trigger decision requires both the correct energy deposition to trigger BETA but not VETO and a certain number of hits, of any level, in the outer and inner TOFs. If this trigger requirement is met, the MTB generates an event ID which allows for merging events between the TOF readout boards (RBs) and the tracker readout electronics. This event ID is then distributed, along with the trigger decision, to the two systems.

Each of the 40 RBs in the TOF receives eight HG waveform signals from the SiPM preamp and the trigger, event ID, and synchronized clock signal from the MTB. The RBs use the Domino Ring Sampler (DRS-4) ASIC. which samples at a rate of 2 GS/s to digitize a full 512 ns history of the waveforms.

There are 20 readout and trigger (RAT) boxes in the TOF; each RAT box has one LTB, two RBs, and a power board (PB) which accepts 24V battery power and provides power to the three other boards in the RAT box as well as the power and bias voltages to 16 SiPM preamps. See Figure 5 for photos of the inside and outside of the RAT boxes.

## 5. Prototype and Testing

There have been three stages of prototyping and testing for the GAPS experiment. The first was the GAPS Functional Prototype (GFP) built at MIT Bates Lab in Boston Massachusetts during the Fall of 2021. The GFP TOF was composed of two parallel panels of twelve 180 cm paddles, separated by approximately a meter and placed over the GFP tracker. For GFP data taking, data from paddles directly on top of each other were read out using the same RB. Applying cuts on the time difference between paddle ends to select only vertical muons passing through the center of both the upper and lower paddles, it was possible to measure timing resolution performance (see

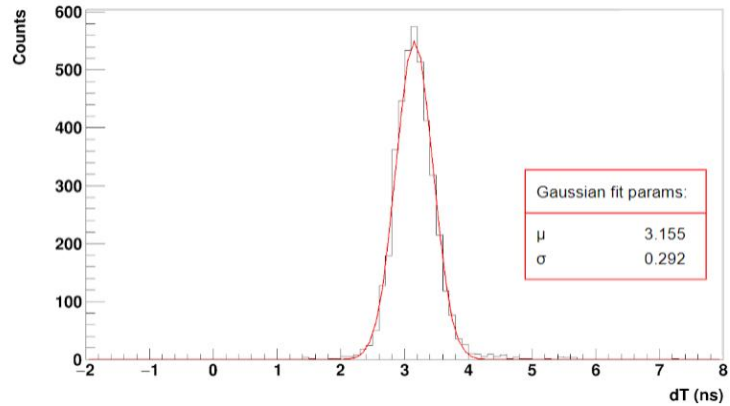

**Figure 6:** Distribution of times of flight for GFP data for which we find a timing resolution of 290ps based on a Gaussian fit.

Figure 6) to 290 ps, which is significantly better than our requirement of 400 ps. This is a significant validation for the design and a good test of hardware and readout systems.

The next phase of testing started in the fall of 2022, with the integration of the entire tracker with the inner TOF at the UC Berkeley Space Sciences Lab. This ground system testing also involved construction of the mechanical structure. Then, in June 2023, thermal and vacuum performance testing was done at National Technical Systems in El Segundo, CA. The TOF was operated for several days at float pressure and at both the hottest and coldest temperatures expected in flight from the thermal model. During these operations, the TOF system performed as expected, with the electronic boards working correctly.

## 6. Conclusion and the Future of GAPS

At the time of this conference, it can be reported that the construction of the TOF components (electronics and hardware) is complete, that results from the GFP indicate that timing resolution goals are met, and that the TVAC testing was very successful. The TOF appears to be robust and ready for a long-duration balloon launch.

GAPS is currently on schedule to launch during the austral summer of 2024-2025. During the Fall of 2023, the payload will undergo further testing at Nevis Labs, Columbia University. In June of 2024, compatibility testing will be performed at NASA facilities in Palestine, TX, in preparation for shipment to Antarctica in the fall of 2024.

## References

[1] Daniel Bauer, James Buckley, Matthew Cahill-Rowley, Randel Cotta, Alex Drlica-Wagner, Jonathan L. Feng, Stefan Funk, et al. Dark matter in the coming decade: Complementary paths to discovery and beyond. *Physics of the Dark Universe*, 7-8:16–23, mar 2015.

[2] Fiorenza Donato. Antimatter from supersymmetric dark matter. In David B. Cline, editor, *Sources and Detection of Dark Matter and Dark Energy in the Universe*, pages 236–243, Berlin, Heidelberg, 2001. Springer Berlin Heidelberg.

[3] Fiorenza Donato, Nicolao Fornengo, and Pierre Salati. Antideuterons as a signature of supersymmetric dark matter. *Physical Review D*, 62(4), jul 2000.

[4] M. Aguilar, L. Ali Cavasonza, B. Alpat, G. Ambrosi, L. Arruda, N. Attig, S. Aupetit, et al. Antiproton flux, antiproton-to-proton flux ratio, and properties of elementary particle fluxes in primary cosmic rays measured with the alpha magnetic spectrometer on the international space station. *Phys. Rev. Lett.*, 117:091103, Aug 2016.

[5] Jan Heisig, Michael Korsmeier, and Martin Wolfgang Winkler. Dark matter or correlated errors: Systematics of the AMS-02 antiproton excess. *Physical Review Research*, 2(4), oct 2020.

[6] Rebecca K. Leane, Seodong Shin, Liang Yang, Govinda Adhikari, Haider Alhazmi, Tsuguo Aramaki, Daniel Baxter, et al. Snowmass2021 cosmic frontier white paper: Puzzling excesses in dark matter searches and how to resolve them, 2022.

[7] R. Munini, E. Vannuccini, M. Boezio, P. von Doetinchem, C. Gerrity, A. Lenni, N. Marcelli, S. Quinn, F. Rogers, J.L. Ryan, A. Stoessl, M. Xiao, N. Saffold, A. Tiberio, and M. Yamatani. The antinucleus annihilation reconstruction algorithm of the GAPS experiment. *Astroparticle Physics*, 133:102640, dec 2021.

[8] Valentina Scotti, Alfonso Boiano, Lorenzo Fabris, Massimo Manghisoni, Giuseppe Osteria, Elisa Riceputi, Francesco Perfetto, Valerio Re, and Gianluigi Zampa. Front-end electronics for the gaps tracker, 2019.

[9] Mengjiao Xiao, Achim Stoessl, Brandon Roach, Cory Gerrity, Ian Bouche, Gabriel Bridges, Philip Von Doetinchem, Charles J. Hailey, Derik Kraych, Anika Katt, Michael Law, Alexander Lowell, Evan Martinez, Kerstin Perez, Maggie Reed, Chelsea Rodriguez, Nathan Saffold, Ceaser Stringfield, Hershel Weiner, and Kelsey Yee. Large-scale detector testing for the gaps si(li) tracker. *IEEE Transactions on Nuclear Science*, pages 1–1, 2023.

[10] H. Fuke, S. Okazaki, H. Ogawa, and Y. Miyazaki. Balloon flight demonstration of an oscillating heat pipe. *Journal of Astronomical Instrumentation*, 06(02):1740006, 2017.

## Full Authors List: GAPS Collaboration

T. Aramaki[1], M. Boezio[2,3], S. E. Boggs[4], V. Bonvicini[2], G. Bridges[5], D. Campana[6], W. W. Craig[7], P. von Doetinchem[8], E. Everson[9], L. Fabris[10], S. Feldman[9], H. Fuke[11], F. Gahbauer[5], C. Gerrity[8], L. Ghislotti[15,16], C. J. Hailey[5], T. Hayashi[5], A. Kawachi[12], M. Kozai[13], P. Lazzaroni[15,16], M. Law[5], A. Lenni[3], A. Lowell[7], M. Manghisoni[15,16], N. Marcelli[18], K. Mizukoshi[27], E. Mocchiutti[2,3], B. Mochizuki[7], S. A. I. Mognet[19], K. Munakata[20], R. Munini[2,3], S. Okazaki[27], J. Olson[22], R. A. Ong[9], G. Osteria[6], K. Perez[5], F. Perfetto[6], S. Quinn[9], V. Re[15,16], E. Riceputi[15,16], B. Roach[23], F. Rogers[7], J. L. Ryan[9], N. Saffold[5], V. Scotti[6,24], Y. Shimizu[25], K. Shutt[7], R. Sparvoli[17,18], A. Stoessl[8], A. Tiberio[26,29], E. Vannuccini[26], M. Xiao[23], M. Yamatani[11], K. Yee[23], T. Yoshida[11], G. Zampa[2], J. Zeng[1], and J. Zweerink[9]

[1]Northeastern University, 360 Huntington Avenue, Boston, MA 02115, USA [2]INFN, Sezione di Trieste, Padriciano 99, I-34149 Trieste, Italy [3]IFPU, Via Beirut 2, I-34014 Trieste, Italy [4]University of California, San Diego, 9500 Gilman Dr., La Jolla, CA 90037, USA [5]Columbia University, 550 West 120th St., New York, NY 10027, USA [6]INFN, Sezione di Napoli, Strada Comunale Cinthia, I-80126 Naples, Italy [7]Space Sciences Laboratory, University of California, Berkeley, 7 Gauss Way, Berkeley, CA 94720, USA [8]University of Hawai'i at Mānoa, 2505 Correa Road, Honolulu, Hawaii 96822, USA [9]University of California, Los Angeles, 475 Portola Plaza, Los Angeles, CA 90095, USA [10]Oak Ridge National Laboratory, 1 Bethel Valley Rd., Oak Ridge, TN 37831, USA [11]Institute of Space and Astronautical Science, Japan Aerospace Exploration Agency (ISAS/JAXA), Sagamihara, Kanagawa 252-5210, Japan [12]Tokai University, Hiratsuka, Kanagawa 259-1292, Japan [13]Polar Environment Data Science Center,Joint Support-Center for Data Science Research,Research Organization of Information and Systems,(PEDSC, ROIS-DS), Tachikawa 190-0014, Japan [15]Università di Bergamo, Viale Marconi 5, I-24044 Dalmine (BG), Italy [16]INFN, Sezione di Pavia, Via Agostino Bassi 6, I-27100 Pavia, Italy [17]INFN, Sezione di Roma "Tor Vergata", Piazzale Aldo Moro 2, I-00133 Rome, Italy [18]Università di Roma "Tor Vergata", Via della Ricerca Scientifica, I-00133 Rome, Italy [19]Pennsylvania State University, 201 Old Main, University Park, PA 16802 USA [20]Shinshu University, Matsumoto, Nagano 390-8621, Japan [22]Heliospace Corporation, 2448 6th St., Berkeley, CA 94710, USA [23]Massachusetts Institute of Technology, 77 Massachusetts Ave., Cambridge, MA 02139, USA [24]Università di Napoli "Federico II", Corso Umberto I 40, I-80138 Naples, Italy [25]Kanagawa University, Yokohama, Kanagawa 221-8686, Japan [26]INFN, Sezione di Firenze, via Sansone 1, I-50019 Sesto Fiorentino, Florence, Italy [27]Research and Development Directorate, Japan Aerospace Exploration Agency (JAXA), 2-1-1 Sengen, Tsukuba 305-8505, Japan [28]Research and Development Directorate, Japan Aerospace Exploration Agency, Sagamihara, Kanagawa 252-5210, Japan [29] INFN, Sezione di Firenze, via Sansone 1, I-50019 Sesto Fiorentino, Florence, Italy

This work is supported in the U.S. by the NASA APRA program (Grant Nos. NNX17AB44G, NNX17AB46G, and NNX17AB47G), in Japan by the JAXA/ISAS Small Science Program FY2017, and in Italy by Istituto Nazionale di Fisica Nucleare (INFN) and the Italian Space Agency (ASI) through the ASI INFN agreement No. 2018-22-HH.0: "Partecipazione italiana al GAPS - General AntiParticle Spectrometer." H. Fuke is supported by JSPS KAKENHI grants (JP17H01136, JP19H05198, and JP22H00147) and Mitsubishi Foundation Research Grant 2019-10038. The contributions of C. Gerrity are supported by NASA under award No. 80NSSC19K1425 of the Future Investigators in NASA Earth and Space Science and Technology (FINESST) program. R. A. Ong receives support from the UCLA Division of Physical Sciences. K. Perez and M. Xiao are supported by Heising-Simons award 2018-0766. Y. Shimizu receives support from JSPS KAKENHI grant JP20K04002 and Sumitomo Foundation Grant No. 180322. M. Yamatani receives support from JSPS KAKENHI grant JP22K14065. K. Yee is supported through the National Science Foundation Graduate Research Fellowship under grant 2141064. S. Feldman is supported through the National Science Foundation Graduate Research Fellowship under grant 2034835.

# The GAPS experiment - a search for light cosmic ray antinuclei

**A. Stoessl**[a,*] **for the GAPS collaboration**

[a]*Departement of Physics and Astronomy,*
*University of Hawaii at Manoa,*
*2505 Correa Rd, Honolulu, HI 96822*

*E-mail:* stoessl@hawaii.edu

The General Anti Particle Spectrometer (GAPS) is a balloon-borne cosmic-ray experiment which is currently in its last phase of construction, undergoing system testing, and scheduled for a long-duration balloon flight from McMurdo Station in the Antarctic in December 2024. Its primary scientific goal is the search for light antinuclei in cosmic rays at kinetic energies below 0.25 GeV/$n$. This energy region is especially of interest for beyond-the-standard model dark matter searches and is still mostly uncharted. Searches for light antimatter nuclei with energies below 0.25 GeV/$n$ are a novel approach to the search for dark matter because wide range of dark matter models proposes annihilation or decay into matter-antimatter pairs. GAPS will yield unprecedented sensitivity to low-energy antideuterons and will measure the low-energy antiproton spectrum with high statistics and precision. To reach the required sensitivity, the GAPS detector incorporates a new approach for antimatter detection, utilizing a tracker with custom, lithium-drifted silicon detectors, designed to measure the X-ray cascade expected from antimatter capture and charged particles from the subsequent annihilation. It also utilizes a fast time-of-flight system, allowing for a high-precision beta measurement. This proceeding highlights GAPS scientific goals.

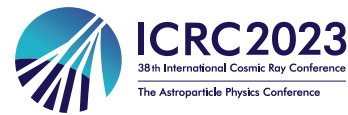

[*]Speaker

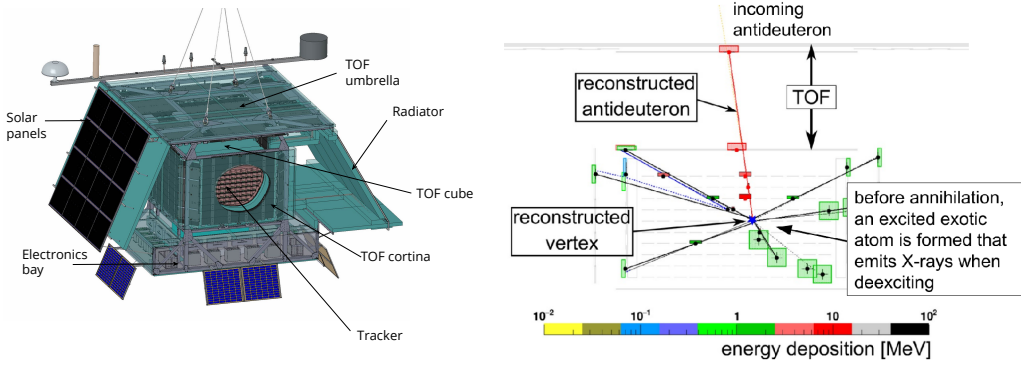

**Fig. 1:** *Left:* A schematic overview of the GAPS instrument: Indicated are the three components of the time of flight (TOF) system as well as the Si(Li) tracker. *Right:* A $\bar{d}$ annihilating in the tracker forming an exotic atom in the process and resulting in a final state with several pions.

## 1 Introduction

In modern physics, the question of the nature of dark matter (DM) is of major importance. Our current understanding of the universe relies on the inclusion of this yet-to-be-identified matter component, currently presenting itself to the observer only through gravitation [1].

Proposed methods for dark matter detection include observing its decay or annihilation products, which might include antinuclei. While antiprotons have been measured in cosmic-ray data [2], conclusive results about their origin are still under debate due to uncertainties in the astrophysical background predictions [3]. Especially secondary $\bar{p}$ fluxes from $pp \Rightarrow p+p+\bar{p}p+...$ at sufficiently high energies are not well understood yet and pose a challenge for uncovering the observed $\bar{p}$ origin. This makes heavier antinuclei such as antideuterons very interesting, since the predicted fluxes for heavier astrophysical antinuclei at low energies $(0.1 - 0.3\,\text{GeV}/n)$ are about 3 orders of magnitude below the predicted fluxes for a variety of dark matter models [3].

The General Antiparticle Spectrometer (GAPS) has been proposed as an indirect dark matter detection experiment optimized to detect these low-energy $(0.1 - 0.3\,\text{GeV}/n)$ cosmic antinuclei using a series of Antarctic long-duration balloon (LDB) flights [4]. The GAPS detection principle does not rely on a magnet and is further described in Section 2. It utilizes the so-called "exotic atom technique" : Similar to a muonic atom, an exotic atom is formed when an antinucleus replaces electrons in the shell of a target atom. This configuration is unstable, and when decaying, the antinucleus will emit a series of characteristic X-rays in the process, which can be used to identify the antinucleus. Ultimately, the antinucleus will annihilate with the nuclei, emitting secondaries. The GAPS scientific payload has been assembled, completed a series of thermal-vacuum tests, and is preparing for its launch in December 2024. Its expected science capabilities for $\bar{p}$, $\bar{d}$ and $\bar{\alpha}$ are outlined in Section 3.

## 2 The GAPS experiment

### 2.1 Instrument overview

The GAPS science instrument is a payload of approximately 2300 kg which will be attached to a balloon suitable for a long-duration balloon flight (LDB) from McMurdo, Antarctica. The first launch is planned for December 2024.
GAPS has two major detector components, a time-of-flight system (TOF) as well as a Si(Li) tracker. To account for the limited power, weight limitations, and reduced bandwidth for data transmission for balloon experiments, GAPS features specifically designed detectors and readout. This includes the utilization of SiPMs for the TOF system, the usage of lightweight styrofoam to hold the individual tracker detectors, a novel passive heat pipe system, and extensive onboard data processing utilizing low-power CPUs. A schematic of the payload is shown in Figure 1.
The TOF system is made of 160 individual scintillator paddles. Each scintillator paddle is read out by 6 SiPMs on each end. The individual paddles are arranged in three main components: A cube, wrapped tightly around the GAPS tracker, a cortina which is a layer around the cube and an umbrella which is a large panel of scintillator paddles on top of the instrument above 1m above the cube.
The umbrella provides a large angular acceptance for the trigger. The two panels are at a distance of about 1m, which does allow for a precise measurement of the incoming primary velocity. Dedicated electronics allow for a tunable trigger based on threshold crossing of the recorded signal in the individual paddles. A so-called master trigger which receives the signal of all paddles over threshold can use pre-programmed paddle combinations, the value of the crossed threshold as well as a measured velocity estimate to provide the global trigger decision. The trigger will reduce the individual, per paddle rate of about 2kHz to a global trigger rate of about 500Hz. Besides providing the trigger, the TOF system records the SiPM waveforms with dedicated electronics utilizing a DRS4 chip. While providing important information about the incoming primary, the TOF is also recording the signals of the secondary particles leaving the tracker after an annihilation. After the waveforms are recorded, they are processed onboard, and a low-power CPU will analyze the event, providing a more precise $\beta$ estimate as well as categorizing it by event quality, allowing it to impose filter criteria. These filter criteria can then be used to decide which events will be transmitted to the ground.

The GAPS tracker has 10 individual tracker layers, 7 of them instrumented with 252 active modules, where a module is an integration unit of 4 individual detectors. The Si(Li) detectors are of custom design and feature a large diameter (4") and are segmented into 8 distinct strips [7]. The detectors are read out by a custom-designed ASIC [6], featuring a wide dynamic range allowing to measure X-rays from $20 - 100$ keV as well as the energy depositions of heavier cosmic-ray (anti)nuclei up to 100 MeV. The Si(Li) detectors can be operated at high temperatures for such detectors of $-40\ °C$. This allows for passive cooling, which is provided by a newly designed oscillating heatpipe system [8].

## 2.2 Detection technique

The instrument is optimized to detect slow-moving ($\beta \approx 0.2 - 0.5$) antinuclei with single or double charge. Signal identification relies on a precise measurement of the particle's velocity, a calorimetric energy measurement, tracking of the annihilation products as well as the measurement of de-excitation X-rays from exotic atoms.

The measurement principle is illustrated in Figure 1. A primary particle, in this case, an antideuteron, passes through the umbrella part of the TOF as well as its upper cube panel, providing a $\beta$ measurement as well as a measurement of the deposited energy along its track. The antideuteron then gets captured by a silicon nucleus in one of the tracker detectors, forming an exotic atom in the process. In these atoms, the incoming antinucleus replaces an electron in the shell before emitting de-excitation X-rays with energies of several 10keV. This exotic atom is decaying almost instantaneously when the antideuteron annihilates with nucleons of the Silicon, resulting in emerging pions and kaons. The decay products are registered by the tracker and TOF and the observed star pattern can be reconstructed to help determine the annihilation vertex, together with the reconstructed incoming primary track. Energy deposition patterns along the tracks together with multiplicity and velocities of secondaries as well as the de-excitation X-rays are used in the identification of the particle species.

## 3 GAPS scientific goals

A confirmed detection of a single cosmic antideuteron would be the first of its kind. Besides being a groundbreaking discovery on its own, antideuterons can help to resolve the antihelium puzzle [3]. Antinuclei have to be the product of their constituents and assuming that the probability for the constituents to coalesce into an antinucleus drops rapidly with the number of constituents, antideuteron formation should be much more likely than antihelium formation. In light of the tentative antihelium candidates reported by the AMS-02 collaboration, this means that an antideuteron measurement by GAPS might help understand the origin of the AMS-02 antihelium candidates. GAPS is not only capable to search in an energy range complimentary to AMS-02, but it also uses a different detection technique with different systematic errors.

The GAPS capabilities have been investigated with large-scale MC simulation productions of various (anti)particle species. The simulation includes modeling of the trigger algorithm, a realistic geometry, including crucial passive detector components close to the Si(Li) detectors, as well as an approximation for the digitization and data processing electronics.

### 3.1 Antiproton spectrum

The antiproton spectrum is projected to be measured with unprecedented statistics in the low-energy range. GAPS is expected to measure about 500 antiprotons per 30-day flight in an energy range of $\approx 0.07 - 0.21$ GeV/$n$ at the top of the atmosphere. Not relying on the more traditional approach of a magnetic spectrometer, this measurement will provide a sizeable sample with orthogonal systematics in an energy range where no previous measurements have been conducted [9].

The analysis considers downgoing primaries with $0.25 < \beta < 0.65$. For particle identification, a series of variables are used which are mainly focussing on the energy deposition pattern along the track, the quality of the event reconstruction as well as the properties of the secondaries from the annihilation. The projected spectrum is shown in Figure 2. It extends previous measurements well into the energy range below 200 MeV/$n$ with very small statistical errors.

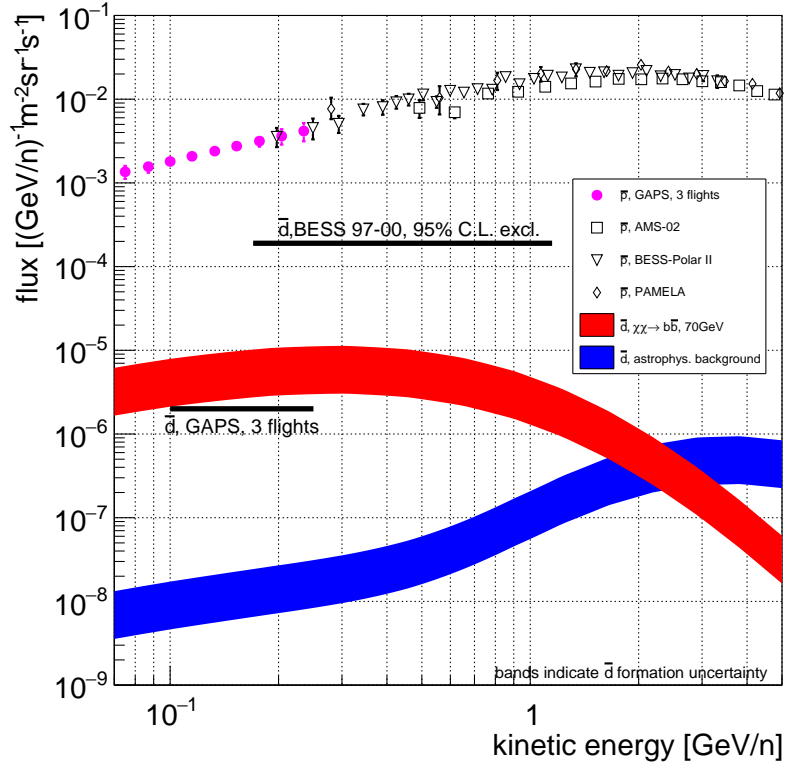

**Fig. 2:** The projected cosmic antiproton spectrum at the top of the atmosphere together with the antideuteron sensitivity. Flux prediction for $\bar{d}$ from a generic WIMP annihilation model as well as astrophysical background predictions are shown as well. Data from BESS, AMS-02 and PAMELA taken from [2, 11, 14, 15, 17]

### 3.2 Antideuteron sensitivity

In a similar way as for the antiproton analysis, the sensitivity for antideuterons has been derived. The sensitivity is constructed in such a way, that a single, confirmed antideuteron event will provide a 3-$\sigma$ discovery. The projected antideuteron sensitivity for GAPS is shown in Figure 2. The expected astrophysical backgrounds in this energy range are below the expected signal from many dark matter models by about 3-4 orders of magnitude. The shown dark matter signal is derived from a 70 GeV WIMP annihilation model. The projected GAPS sensitivty improves over the BESS limit [11] by over two orders of magnitude. The model-independent approach to search for antideuterons allows constraining a wide variety of DM models which emerged in recent years with a large variety of antideuteron production mechanisms, such as gravitino decay, extra dimensions, and dark photons. An overview of such models can be found in [3].

### 3.3 Antihelium sensitivity

Compared to antideuterons, antihelium nuclei will deposit more energy along their tracks, due to having twice the charge. The GAPS tracker is designed with a large dynamic range, and thus allows to precisely resolve the energy depositions along the primary track. Also, the higher multiplicity of the secondaries allows for good identification of the resulting annihilation star pattern. The antihelium analysis is conducted similarly to the antideuteron analysis [13]. The expected low astrophysical antihelium backgrounds allow for a strong sensitivity on the order of current dark matter models, as it is shown in Figure 3.

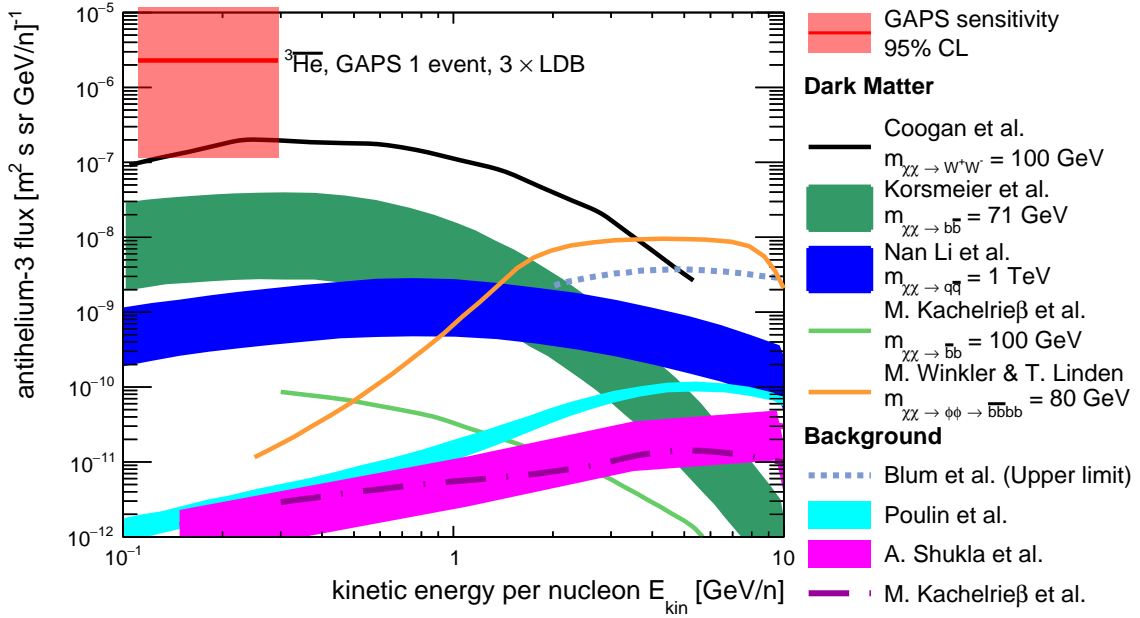

**Fig. 3:** The solid red line shows the single event sensitivity of GAPS to antihelium-3 nuclei (95% confidence level) for three LDB flights of 35 days each. The red box indicates the upper and lower bounds of the 95% confidence level. Also shown is the antihelium-3 flux predicted by a variety of dark matter [18–22] and standard astrophysical background [23–25] models. For theoretical predictions, the error bands illustrate uncertainties in the coalescence momentum but also include propagation uncertainties.

# 4 Conclusion

The GAPS experiment has the potential to expand our understanding of low-energy antinuclei in cosmic rays significantly: Its large acceptance, as well as its novel detection principle, providing sensitivity to antideuteron and antihelium fluxes, which are predicted by current dark matter models and extend the antiproton spectrum with a high statistics spectrum towards low energies. Its unique approach to antiparticle identification, using an exotic atom technique, will provide complimentary systematics to current experiments searching for cosmic antinuclei. The design phase of GAPS has been concluded, and the instrument has been integrated. GAPS has recently undergone extensive thermal-vacuum testing and is currently in its last integration and testing phase before its first flight in December 2024.

# References

[1] K. Freese, *Proceedings of 14th Marcel Grossman Meeting* **MG14 (2015)**

[2] M. Aguilar et al. *Physics Review Letters* **117** (2016) 091103.

[3] P.v. Doetinchem et al., *Journal of Cosmology and Astroparticle Physics* **2020** (2020) 035.

[4] T. Aramaki et al. *Astropart.Phys.* **74** (2016) 6-13

[5] S. Feldman et al., *This proceedings* .

[6] M. Manghisoni et al. *IEEE Trans.Nucl.Sci* **68** (2021) 11

[7] K. Perez et al., *Nucl. Instrum. Meth. A* **905** (2018) 12 [1807.07912].

[8] S. Okazaki et al., *Applied Thermal Engineering* **141** (2018) 20 .

[9] F. Rogers et al., *Astropart. Phys.* **145** (2023) 102791

[10] M. Xiao, A. Stoessl, B. Roach, C. Gerrity, I. Bouche et al. e-Print: 2305.00283 [physics.ins-det] **(2023)** .

[11] K. Abe et al., *Phys. Rev. Lett.* **108** (2012) 131301.

[12] D. Gomez et al *PHys Rev. D* **98** (2018) 023012.

[13] N. Saffold, T. Aramaki, R. Bird, M. Boezio, S. Boggs, V. Bonvicini et al., *Astroparticle Physics* **130** (2021) 102580.

[14] M. Aguilar et al. *Physics Reports* **894** (2021) 1–116.

[15] S. Orito et al *Physical Review Letters* **84** (2000) 1078–1081.

[16] K. Abe et al *Physics Review Letters* **108** (2012) 051102.

[17] O. Adriani et al. *Soviet Journal of Experimental and Theoretical Physics Letters* **96** (2013) 621–627.

[18] M. Korsmeier, F. Donato and N. Fornengo, *Physical Review D* **97** (2018) 103011.

[19] A. Coogan and S. Profumo, *Physical Review D* **96** (2017) 083020.

[20] K. Blum et al., *Physical Review D* **96** (2017) .

[21] Y.-C. Ding et al., *Journal of Cosmology and Astroparticle Physics* **2019** (2019) 004.

[22] M.W. Winkler and T. Linden, *Physical Review Letters* **126** (2021) .

[23] V. Poulin et al., *Physical Review D* **99** (2019) 023016 [1808.08961].

[24] M. Kachelrieß, S. Ostapchenko and J. Tjemsland, *Journal of Cosmology and Astroparticle Physics* **2020** (2020) 048.

[25] A. Shukla et al., *Physical Review D* **102** (2020) 063004.

# Full Authors List: GAPS Collaboration

T. Aramaki[1], M. Boezio[2,3], S. E. Boggs[4], V. Bonvicini[2], G. Bridges[5], D. Campana[6], W. W. Craig[7], P. von Doetinchem[8], E. Everson[9], L. Fabris[10], S. Feldman[9], H. Fuke[11], F. Gahbauer[5], C. Gerrity[8], L. Ghislotti[15,16], C. J. Hailey[5], T. Hayashi[5], A. Kawachi[12], M. Kozai[13], P. Lazzaroni[15,16], M. Law[5], A. Lenni[3], A. Lowell[7], M. Manghisoni[15,16], N. Marcelli[18], K. Mizukoshi[27], E. Mocchiutti[2,3], B. Mochizuki[7], S. A. I. Mognet[19], K. Munakata[20], R. Munini[2,3], S. Okazaki[27], J. Olson[22], R. A. Ong[9], G. Osteria[6], K. Perez[5], F. Perfetto[6], S. Quinn[9], V. Re[15,16], E. Riceputi[15,16], B. Roach[23], F. Rogers[7], J. L. Ryan[9], N. Saffold[5], V. Scotti[6,24], Y. Shimizu[25], K. Shutt[7], R. Sparvoli[17,18], A. Stoessl[8], A. Tiberio[26], E. Vannuccini[26], M. Xiao[23], M. Yamatani[11], K. Yee[23], T. Yoshida[11], G. Zampa[2], J. Zeng[1], and J. Zweerink[9]

[1]Northeastern University, 360 Huntington Avenue, Boston, MA 02115, USA [2]INFN, Sezione di Trieste, Padriciano 99, I-34149 Trieste, Italy [3]IFPU, Via Beirut 2, I-34014 Trieste, Italy [4]University of California, San Diego, 9500 Gilman Dr., La Jolla, CA 90037, USA [5]Columbia University, 550 West 120th St., New York, NY 10027, USA [6]INFN, Sezione di Napoli, Strada Comunale Cinthia, I-80126 Naples, Italy [7]Space Sciences Laboratory, University of California, Berkeley, 7 Gauss Way, Berkeley, CA 94720, USA [8]University of Hawai'i at Mānoa, 2505 Correa Road, Honolulu, Hawaii 96822, USA [9]University of California, Los Angeles, 475 Portola Plaza, Los Angeles, CA 90095, USA [10]Oak Ridge National Laboratory, 1 Bethel Valley Rd., Oak Ridge, TN 37831, USA [11]Institute of Space and Astronautical Science, Japan Aerospace Exploration Agency (ISAS/JAXA), Sagamihara, Kanagawa 252-5210, Japan [12]Tokai University, Hiratsuka, Kanagawa 259-1292, Japan [13]Polar Environment Data Science Center,Joint Support-Center for Data Science Research,Research Organization of Information and Systems,(PEDSC, ROIS-DS), Tachikawa 190-0014, Japan [15]INFN, Sezione di Pavia, Via Agostino Bassi 6, I-27100 Pavia, Italy [16]Università di Bergamo, Viale Marconi 5, I-24044 Dalmine (BG), Italy [17]INFN, Sezione di Roma "Tor Vergata", Piazzale Aldo Moro 2, I-00133 Rome, Italy [18]Università di Roma "Tor Vergata", Via della Ricerca Scientifica, I-00133 Rome, Italy [19]Pennsylvania State University, 201 Old Main, University Park, PA 16802 USA [20]Shinshu University, Matsumoto, Nagano 390-8621, Japan [22]Heliospace Corporation, 2448 6th St., Berkeley, CA 94710, USA [23]Massachusetts Institute of Technology, 77 Massachusetts Ave., Cambridge, MA 02139, USA [24]Università di Napoli "Federico II", Corso Umberto I 40, I-80138 Naples, Italy [25]Kanagawa University, Yokohama, Kanagawa 221-8686, Japan [26]INFN, Sezione di Firenze, via Sansone 1, I-50019 Sesto Fiorentino, Florence, Italy [27]Research and Development Directorate, Japan Aerospace Exploration Agency (JAXA), 2-1-1 Sengen, Tsukuba 305-8505, Japan [28]Research and Development Directorate, Japan Aerospace Exploration Agency, Sagamihara, Kanagawa 252-5210, Japan

Web version: gaps1.astro.ucla.edu/gaps/authors/

# Integration and Calibration of the GAPS Antarctic Balloon Payload

**Riccardo Munini**[a,b,*] **and Field Rogers**[b] **for the GAPS collaboration**

[a]*INFN, Sezione di Trieste, I-34149 Trieste, Italy*
[b]*IFPU, I-34014 Trieste, Italy*
[c]*SSL UC Berkeley*

E-mail: riccardo.munini@ts.infn.it

With its inaugural Antarctic long-duration balloon mission in December 2024, the General Antiparticle Spectrometer (GAPS) will become the first experiment optimized to detect cosmic-ray antinuclei below 0.25 GeV/n. Detection of a single antideuteron in this energy range would be a smoking-gun signature of new physics, such as dark matter. The GAPS program will also provide a precision antiproton spectrum in a previously unprobed low-energy range, as well as leading sensitivity to antihelium-3. This new parameter space is accessible thanks to a novel particle identification method based on exotic atom formation, de-excitation, and annihilation. The method provides a unique signature for the negatively-charged antinuclei, facilitating excellent rejection of the positive-nucleus background, and does not require a magnet, enabling a large sensitive area for rare events. The GAPS instrument is designed to provide excellent discrimination power for rare events within the power and mass constraints of a long-duration balloon. The time-of-flight, composed of 160 scintillator paddles, provides the system trigger as well as the GAPS energy scale. The 2.5 m$^3$ tracker volume is instrumented with 10-cm-diameter silicon sensors, which serve as the target, X-ray spectrometer, and particle tracker. Together, a large-area radiator and an integrated oscillating heat pipe system cool the payload without a bulky cryostat. This contribution reports the integration and calibration of the GAPS science payload, including the performance of the sensitive detector subsystems, the cooling system, the power distribution, and data acquisition and onboard event processing.

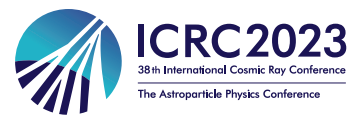

[*]Speaker

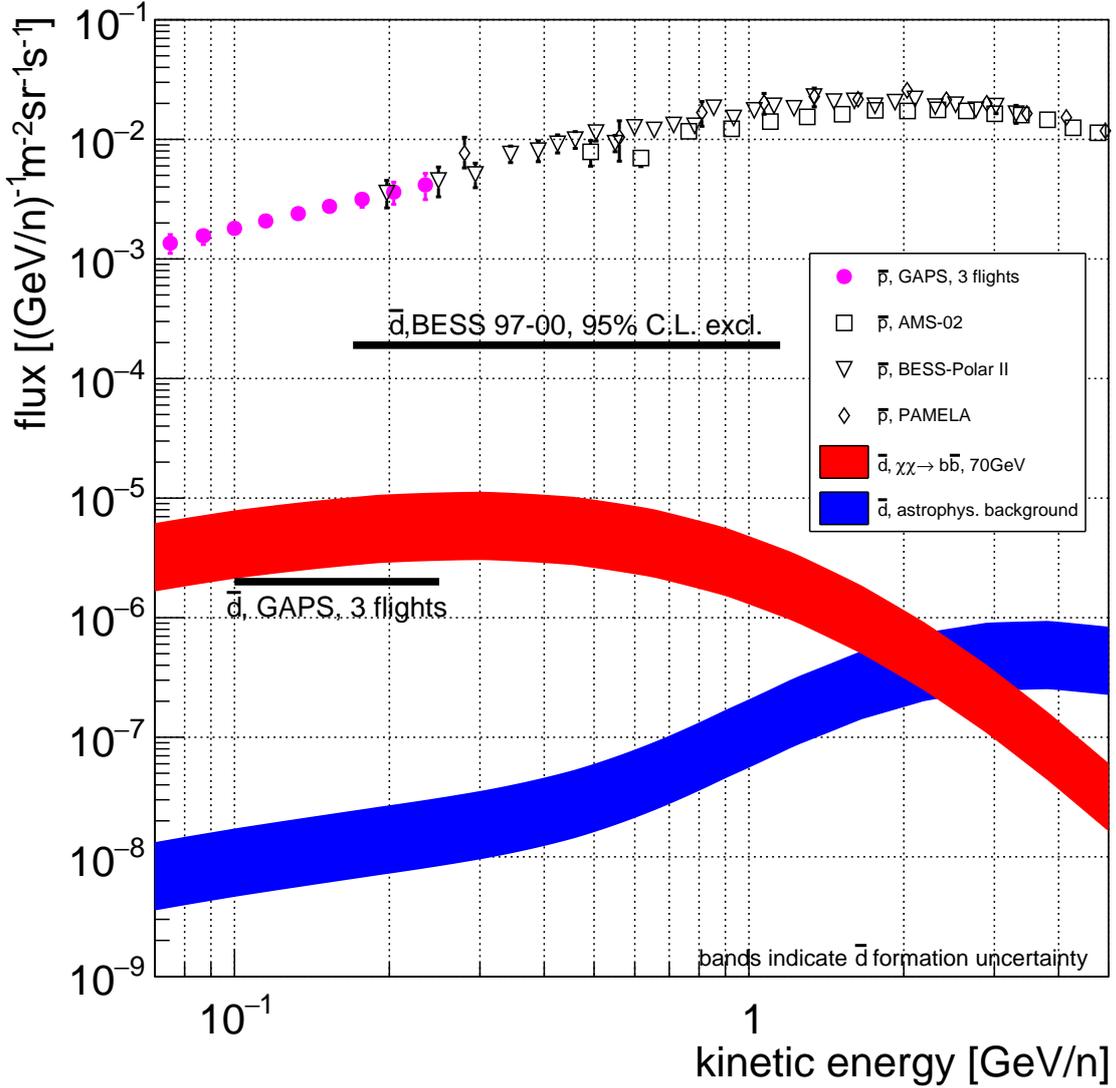

**Figure 1:** The expected sensitivity of the GAPS Experiment to cosmic-ray antideuterons is compared to the predicted flux from a sample DM model (red) and predicted background from known astrophysics astrophysical backgrounds (blue, green) as well as limits by the BESS Experiment. Meanwhile, the expected GAPS sensitivity to antiprotons (orange) will result in a precision spectrum extending to lower energies than past measurements.

## 1. The GAPS science

The General Antiparticle Spectrometer (GAPS) [1–6] is the first experiment optimized to measure low-energy cosmic-ray antideuterons. As illustrated in Figure 1, low-energy cosmic-ray antideuterons constitute a dark matter (DM) probe that is uniquely free of astrophysical background. While antideuterons are expected from a range of DM models, production via known astrophysical processes is kinematically suppressed, so detection of low-energy antideuterons would indicate new physics [7].

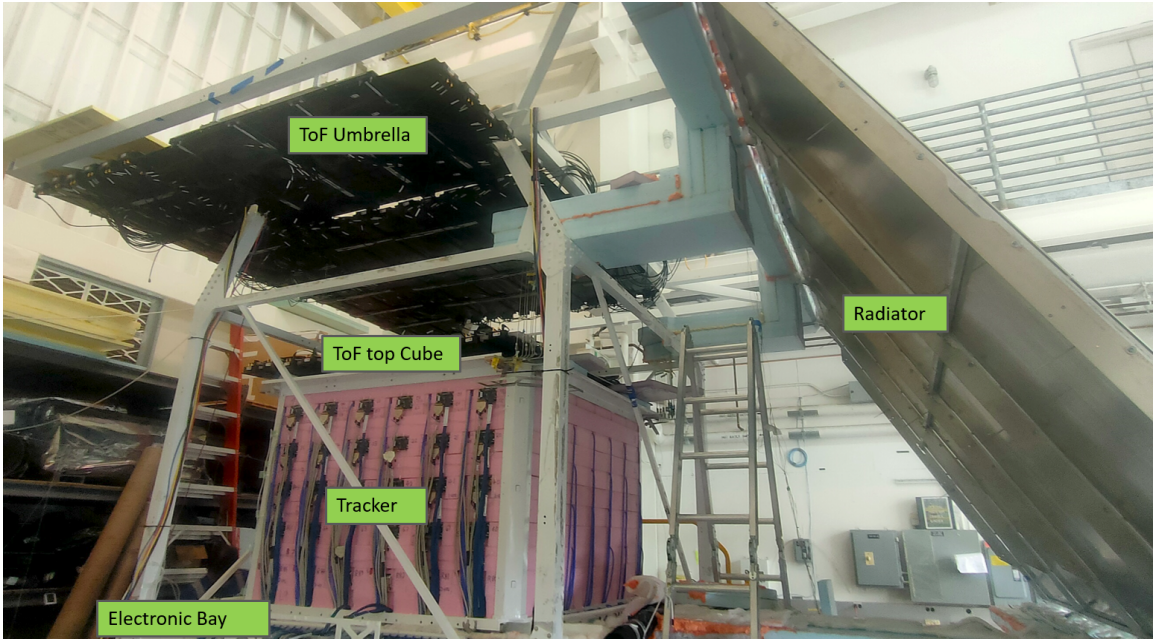

**Figure 2:** The GAPS experiment integrated at SSL. The 10-layer tracker is the core of the apparatus. The TOF appears partially integrated with the plastic scintillators forming the top Cube and the Umbrella. The radiatior of the passive cooling system is shown on the right.

The GAPS detector is designed for the kinetic energies below 0.25 GeV/n, where the signal-to-background is maximized for antideuterons from DM. GAPS will operate from an Antarctic long-duration balloon, which provides flights of about 35 days at 37 km altitude, and a latitude that minimizes geomagnetic effects. The first flight is expected for 2024. A novel exotic-atom-based detection technique provides identification power for antideuterons and a large sensitive area for rare events within the constraints of a balloon payload. Antinuclei are identified using a velocity measurement in combination with exotic atom signatures. GAPS is expected to produce a high-statistics measurement of the antiproton flux in the kinetic energy range of 0.08-0.25 GeV/n. It will also perform rare event searches resulting in leading sensitivity to low-energy antideuteron and antihelium-3 cosmic-ray fluxes. Figure 1 shows the expected antideuteron sensitivity assuming three flights of 30 days each.

## 2. The GAPS experiment

Figure 2 illustrates the integration of the GAPS science payload. The instrument consists of a TOF, which provides the velocity measurement and the trigger. It surrounds the tracker which serves as target and X-ray spectrometer. The TOF, composed of $160 \times 0.63 \times (108 - 180)$ cm$^3$ scintillator paddles, provides timing resolution < 400 ps using silicon-photomultiplier (SiPM) sensors and custom readout electronics. This subsystem consists of an inner Cube which encloses the tracker volume and an Umbrella panel positioned 90 cm above the Cube. In addition a Cortina surrounds the vertical sides of the Cube with a spacing of 30 cm. Figure 3 shows a close view of the Umbrella and top Cube Time of Flight integrated at Space Sciences Laboratory (SSL) in Berkeley as well as a single plastic bar with the SiPm attached at its end. The 10-layer tracker is composed of 1440 8-strip, 10-cm diamter, 2.5 mm thick lithium drifted silicon (Si(Li)) sensors [8] instrumentng a 2.5m$^3$ volume. Seven planes are read out with custom application specific integrated circuits [9]. The other 3 planes provide additional target mass and thermal balance. Each plane is composed of 6 × 6 modules, with one module consisting of four Si(Li) wafer each. Figure 3 shows the top layer of the tracker with a close view of one module with the four Si(Li) cylinders. The tracker is cooled in flight to -40°C using an oscillating heat-pipe (OHP) system [10] coupled to a radiator. In flight, the radiator points away from the sun, while during ground calibrations, it is coupled to a metal plate cooled by circulating -80°C methanol. The tracker is designed to provide <4 keV resolution for 20 - 100 keV X-rays as well as a dynamic range up to 100 MeV. This contribution discusses the integration of the GAPS tracker, TOF, thermal, power, and flight software subsystems. It additionally reports on the calibration of the detector subsystems and of the instrument as a whole.

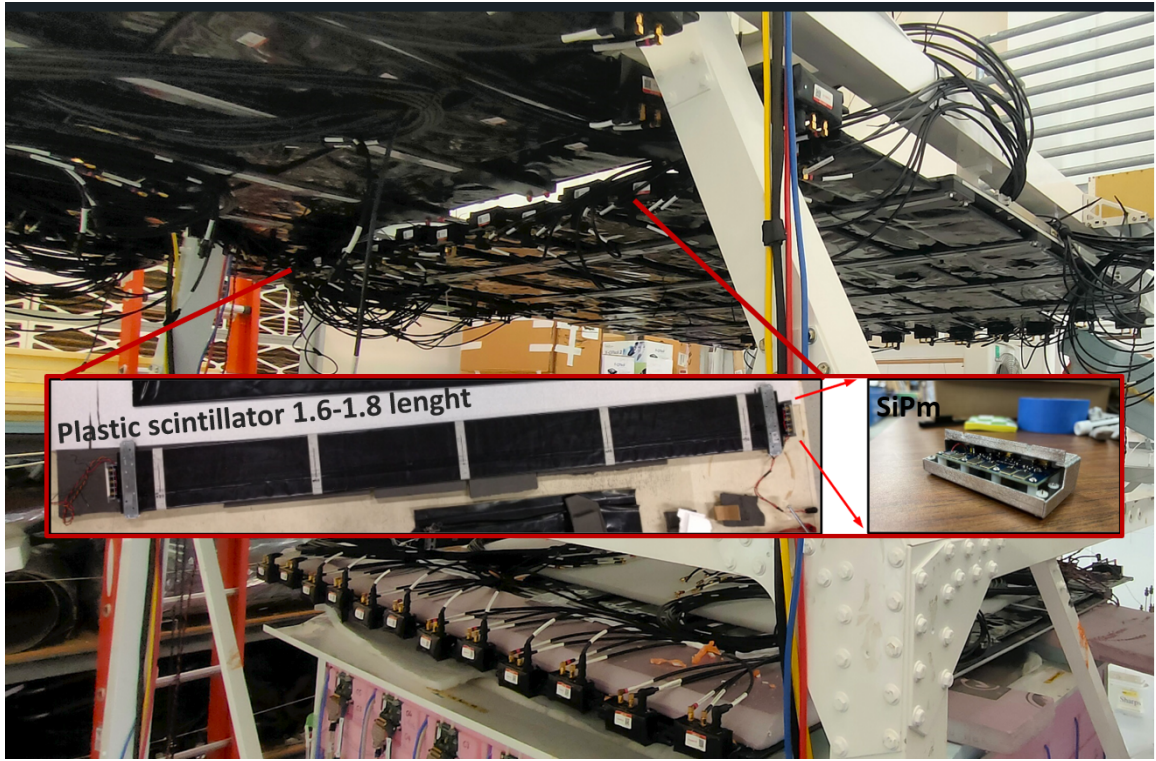

**Figure 3:** The time of flight system composed of bars of plastic scintillators read out by SiPm.

## 3. The GAPS integration timeline

- Spring-Summer 2022: GAPS integration started at Bates Massachusetts Institute of Technology (MIT) laboratories with the mechanical structure assembling. Also the tracker integration started at MIT with a total of five planes integrated during this period.

- September 2022-February 2023: The instrument was moved to the SSL, University of California at Berkeley, where the remaining tracker planes were integrated. The tracker was

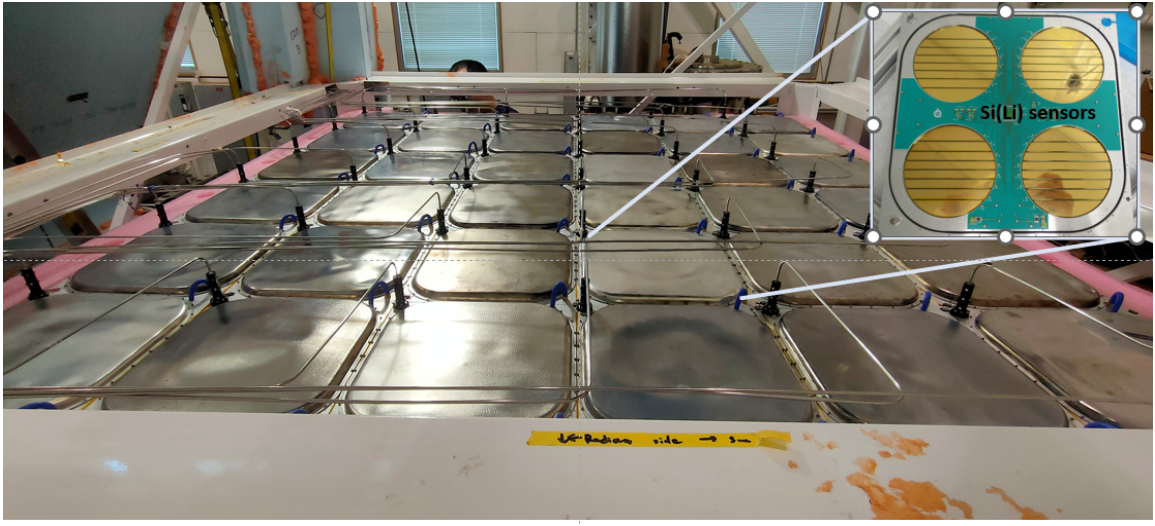

**Figure 4:** The top layer of the GAPS tracker. Each layer is composed of 6×6 modules made of 4 Si(Li) wafers divided into 8 strips (see zoomed pictures on the top right). The tubes attached to each module comprise the cooling system and ensure a nominal operating temperature between -40°and -45 °.

- calibrated using cosmic muons, radioactive sources, and the self-calibration modes of the ASIC.

- Late 2022: the cooling system was fully integrated and tested. The thermal performance on the ground was monitored using thermocouples positioned throughout the tracker. The nominal performance was reached with the tracker modules all at approximately -30°C.

- April-May 2023: the top Cube and part of the Umbrella ToF are integrated and tested. The TOF timing response, energy scale, and velocity calculation are calibrated using cosmic muons. The tracker was also tested with muon acquisition using the ToF trigger. The system interfaces and integrated performance are also calibrated using cosmic muons.

- June 2023: GAPS was shipped to the National Technical System (NTS) facility in Los Angeles for thermal vacuum test (TVAC). The TVAC validated the comprehensive thermal model for the instrument. Both the overall instrument temperatures from the coldest to the hottest temperatures expected, as well as instrument subsystems, operated consistent with this model. The GAPS instrument had full functionality, and muons were recorded on all detector modules except for one that had a high voltage issue, which will be repaired.

- July 2023: the instrument was moved to Columbia Nevis Laboratory.

## 4. Future integration time-line

- Fall 2023: the instrument will be reassembled at Columbia laboratories in Nevis, New York.

- Winter 2023 - Spring 2024: extensive muon data taking for tracker calibration and TOF system operation will be performed. Tracker Si(Li) will be calibrated with X-ray sources.

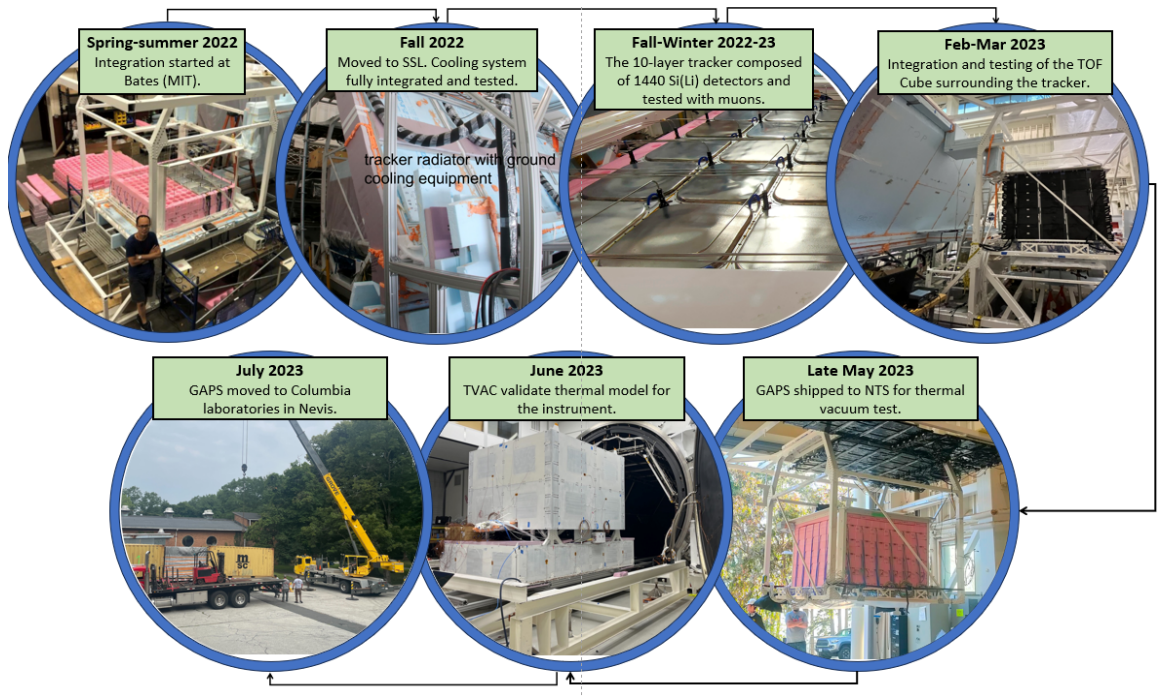

**Figure 5:** Timeline for the integration of the GAPS instrument to date.

- June 2024: the instrument will be moved to NASA facilities in Palestine, Texas for integration with the ballooncraft and compatibility and hang test.
- Fall 2024: the detector will be shipped to the McMurdo station where GAPS will be tested before the launch.
- December 2024: launch for the first 30 duration balloon flight.

# References

[1] K. Mori et al., deexcitation of exotic atoms," ApJ vol. 566, 2002

[2] C.J. Hailey, New J Phys., vol. 11, 2009

[3] T. Aramaki et al., Astropart. Phys., vol. 74, 6–13, 2016

[4] T. Aramaki et al., Astropart. Phys., vol. 59, 12–17, 2014

[5] N. Saffold et al., Astropart. Phys., vol. 130, 102580, 2021

[6] F. Rogers et al., Astropart. Phys., vol. 145, 2023

[7] N. Fornengo et al., JCAP, vol. 2013, 2013.

[8] F. Rogers et al., JINST, vol. 14, 2019

[9] M. Manghisoni et al., In Proceedings of IEEE NSS/MIC, 2021

[10] H. Fuke et al., In Proceedings of IEEE NSS/MIC, 2019

## Full Authors List: GAPS

T. Aramaki[1], M. Boezio[2,3], S. E. Boggs[4], V. Bonvicini[2], G. Bridges[5], D. Campana[6], W. W. Craig[7], P. von Doetinchem[8], E. Everson[9], L. Fabris[10], S. Feldman[9], H. Fuke[11], F. Gahbauer[5], C. Gerrity[8], L. Ghislotti[15,16], C. J. Hailey[5], T. Hayashi[5], A. Kawachi[12], M. Kozai[13], P. Lazzaroni[15,16], M. Law[5], A. Lenni[3], A. Lowell[7], M. Manghisoni[15,16], N. Marcelli[18], K. Mizukoshi[27], E. Mocchiutti[2,3], B. Mochizuki[7], S. A. I. Mognet[19], K. Munakata[20], R. Munini[2,3], S. Okazaki[27], J. Olson[22], R. A. Ong[9], G. Osteria[6], K. Perez[5], F. Perfetto[6], S. Quinn[9], V. Re[15,16], E. Riceputi[15,16], B. Roach[23], F. Rogers[7], J. L. Ryan[9], N. Saffold[5], V. Scotti[6,24], Y. Shimizu[25], K. Shutt[7], R. Sparvoli[17,18], A. Stoessl[8], A. Tiberio[26,29], E. Vannuccini[26], M. Xiao[23], M. Yamatani[11], K. Yee[23], T. Yoshida[11], G. Zampa[2], J. Zeng[1], and J. Zweerink[9]

[1]Northeastern University, 360 Huntington Avenue, Boston, MA 02115, USA [2]INFN, Sezione di Trieste, Padriciano 99, I-34149 Trieste, Italy [3]IFPU, Via Beirut 2, I-34014 Trieste, Italy [4]University of California, San Diego, 9500 Gilman Dr., La Jolla, CA 90037, USA [5]Columbia University, 550 West 120th St., New York, NY 10027, USA [6]INFN, Sezione di Napoli, Strada Comunale Cinthia, I-80126 Naples, Italy [7]Space Sciences Laboratory, University of California, Berkeley, 7 Gauss Way, Berkeley, CA 94720, USA [8]University of Hawai'i at Mānoa, 2505 Correa Road, Honolulu, Hawaii 96822, USA [9]University of California, Los Angeles, 475 Portola Plaza, Los Angeles, CA 90095, USA [10]Oak Ridge National Laboratory, 1 Bethel Valley Rd., Oak Ridge, TN 37831, USA [11]Institute of Space and Astronautical Science, Japan Aerospace Exploration Agency (ISAS/JAXA), Sagamihara, Kanagawa 252-5210, Japan [12]Tokai University, Hiratsuka, Kanagawa 259-1292, Japan [13]Polar Environment Data Science Center,Joint Support-Center for Data Science Research,Research Organization of Information and Systems,(PEDSC, ROIS-DS), Tachikawa 190-0014, Japan [15]Università di Bergamo, Viale Marconi 5, I-24044 Dalmine (BG), Italy [16]INFN, Sezione di Pavia, Via Agostino Bassi 6, I-27100 Pavia, Italy [17]INFN, Sezione di Roma "Tor Vergata", Piazzale Aldo Moro 2, I-00133 Rome, Italy [18]Università di Roma "Tor Vergata", Via della Ricerca Scientifica, I-00133 Rome, Italy [19]Pennsylvania State University, 201 Old Main, University Park, PA 16802 USA [20]Shinshu University, Matsumoto, Nagano 390-8621, Japan [22]Heliospace Corporation, 2448 6th St., Berkeley, CA 94710, USA [23]Massachusetts Institute of Technology, 77 Massachusetts Ave., Cambridge, MA 02139, USA [24]Università di Napoli "Federico II", Corso Umberto I 40, I-80138 Naples, Italy [25]Kanagawa University, Yokohama, Kanagawa 221-8686, Japan [26]INFN, Sezione di Firenze, via Sansone 1, I-50019 Sesto Fiorentino, Florence, Italy [27]Research and Development Directorate, Japan Aerospace Exploration Agency (JAXA), 2-1-1 Sengen, Tsukuba 305-8505, Japan [28]Research and Development Directorate, Japan Aerospace Exploration Agency, Sagamihara, Kanagawa 252-5210, Japan [29] Universita' degli studi di Firenze, via Sansone 1, I-50019 Sesto Fiorentino, Florence, Italy

# The identification of the cosmic-ray light nuclei with the GAPS experiment

**Riccardo Munini**$^{a,b,*}$ **and Alex Lenni**$^b$ **for the GAPS collaboration**

$^a$*INFN, Sezione di Trieste, I-34149 Trieste, Italy*
$^b$*IFPU, I-34014 Trieste, Italy*

 *E-mail:* riccardo.munini@ts.infn.it, alex.lenni@sissa.it

GAPS (General AntiParticle Spectrometer) is a balloon-borne, large-acceptance experiment designed to detect low-energy (< 0.25 GeV/*n*) cosmic-ray antinuclei during three ~35-day Antarctic flights, with the first of these planned for the 2024-2025 austral summer. The GAPS experiment, currently in preparation for the first flight, consists of a tracker equipped with large-area, lithium-drifted silicon detectors, surrounded by a large-acceptance time-of-flight system made of plastic scintillators. This design has been optimized to perform a novel antiparticle identification technique based on an antinucleus capture and the subsequent exotic atom formation and decay, allowing for large geometrical acceptance because no magnet is required. Although detecting cosmic-ray antinuclei as an indirect dark-matter signature is the primary goal of GAPS, many low-energy cosmic-ray nuclei will also be recorded. Nuclei do not form exotic atoms in the GAPS detectors, and their detection is based on the measurements of the ionization energy depositions, evaluation of the kinetic energy, and the stopping depth relative to the measured velocity. An algorithm was developed to fit the slow-down of particles and antiparticles tracked inside GAPS. The quantities fitted by this algorithm, together with the measured velocity and energy deposition information, allow the identification of protons, deuterons, and helium nuclei and the measurement of their spectra in the low-energy range (< 0.25 GeV/n). The results of this analysis, based on detailed Monte Carlo simulation studies, will be presented in this contribution.

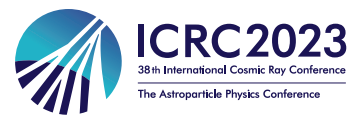

*Speaker

## 1. The GAPS experiment

GAPS (General AntiParticle Spectrometer) is a balloon-borne experiment designed to measure low-energy (< 0.25 GeV/*n*) cosmic-ray antinuclei, especially antiprotons ($\overline{p}$) and antideuterons ($\overline{d}$), during a series of three long-duration flights at an altitude of ∼37 km (∼7 g cm$^{-2}$ of overhead atmosphere) in Antarctica.

The GAPS experiment includes a 10 layer tracker equipped with large-area lithium-drifted silicon (Si(Li)) detectors, surrounded by a Time-of-Flight (TOF) system of thin plastic scintillators (see Figure 1). This design has been optimized to perform a novel antiparticle-identification technique based on antinucleus capture and the subsequent exotic atom formation and decay [1] through a series of atomic transitions emitting characteristic X-rays. Finally, the annihilation of the antinucleus with the atomic nucleus produces a nuclear star of several secondary particles, mainly pions and protons, originating from a common vertex. The instrumental design allows for a large geometrical acceptance since no magnet is required on the order of ∼10 m$^2$ sr in the energy range [0.05; 0.25] GeV/*n*. For more technical and scientific details on the GAPS experiment see [2–7].

The GAPS experiment is scheduled for its first long balloon campaign for the 2024-2025 Austral summer from the McMurdo station in Antarctica. The polar location is ideal for studies of low-energy cosmic rays because of the very low geomagnetic cut-off.

## 2. Light-nuclei with GAPS

Although the GAPS experiment is specifically designed for the detection of cosmic-ray antinuclei, the measurement of low-energy cosmic-ray nuclei is possible and well-motivated for several reasons:

- The study of nucleus events with non-interacting single tracks allows to calibrate the instrument and to gain insight into the systematic uncertainties.

- A high-statistics test of the identification capabilities with protons and deuterons is preparatory to the $\overline{p}/\overline{d}$ analysis. Additionally, the $\overline{p}/\overline{d}$ identifications will benefit from the information of the annihilation (number of secondaries, X-rays, etc.), which will result in even higher rejection power.

- Nuclei measurements will improve the understanding of the propagation of cosmic-ray light nuclei, especially in the Heliosphere which heavily affects the propagation at low energies. Thanks to the foreseen three flights, the nuclei fluxes will be measured during different conditions of solar activity, allowing to test and calibrate cosmic ray solar modulation models [8].

- Moreover, models for secondary production in the atmosphere will be tested with a precision of a few percents.

## 3. Light-nuclei identification

The identification performance of light nuclei was studied with simulated samples of protons, deuterons, and helium nuclei. These data were generated by a GEANT4-based simulation software developed to reproduce the GAPS apparatus and the particle interactions inside it. A custom algorithm [9], developed to reconstruct the topology of the annihilation stars generated by the antinuclei stopping inside the GAPS experiment, was used in this case to reconstruct the primary tracks of non-interacting nucleus. In contrast to antinuclei, the stopping nuclei inside the GAPS instrument do not form exotic atoms in the target material.

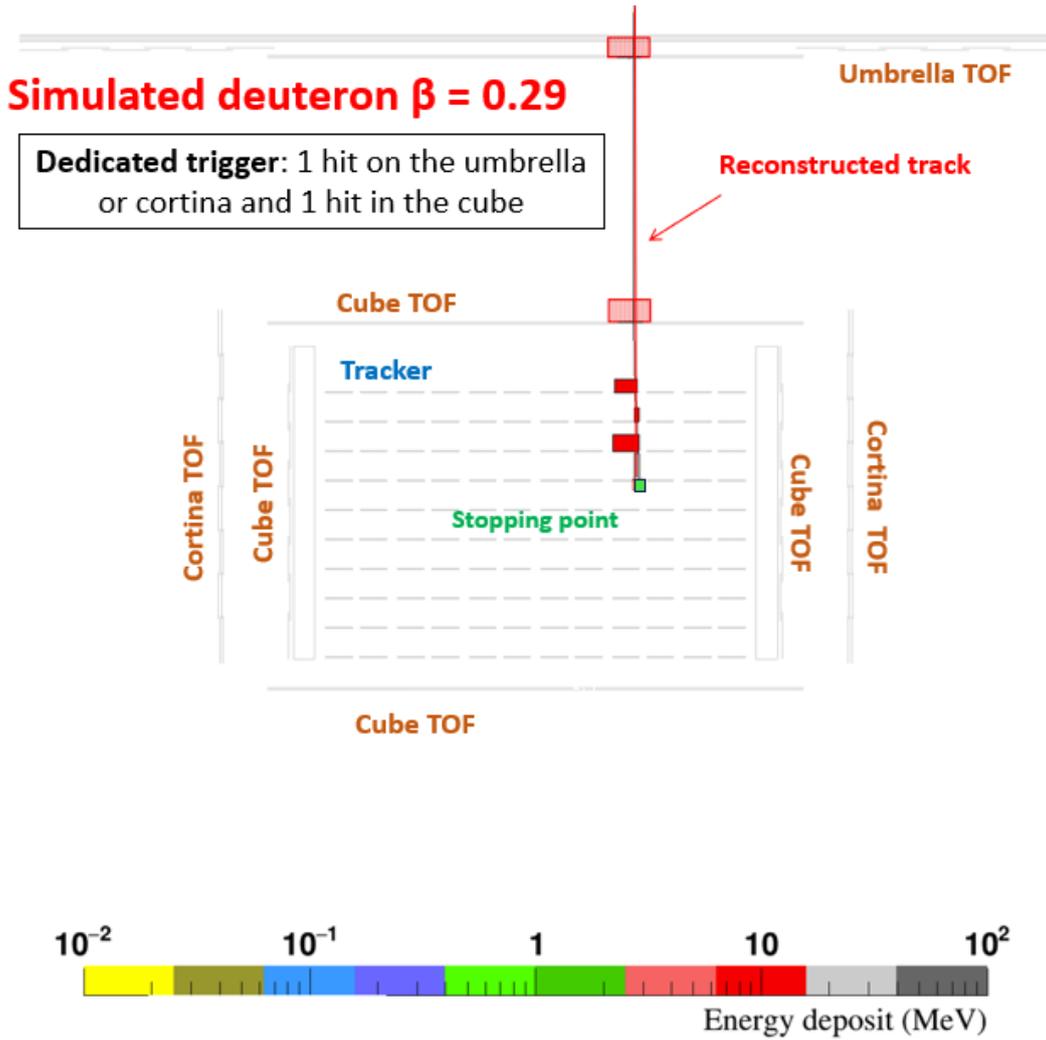

**Figure 1:** Simulated deuteron with a kinetic energy of 90 MeV/n displayed in the XZ view. The event was selected with the minimum bias trigger requiring one energy deposit in the umbrella ToF and one energy deposition in the upper cube ToF. The black line represents the Monte Carlo trajectory while the red line represented the reconstructed trajectory based on the hit associated with the track.

A minimum bias trigger was required to select non-interacting nuclei. The trigger requires a coincidence between one hit in the Umbrella or Cortina ToF and one hit in the Cube ToF. This trigger, in flight, would result in rates greater than 50 kHz. For this reason, the minimum bias trigger will be prescaled with a trigger specifically designed for annihilating events which will result in a rate of less than 100 Hz.

In addition, only events crossing the Umbrella and top Cube downward were selected to reduce the component of secondary produced in the atmosphere, which is more relevant for very inclined events. A dedicated analysis of spectra measured at different angles to perform tomography of the atmosphere. Figure 1 shows an example of a simulated deuteron with a kinetic energy of 90 MeV/$n$ stopping in the tracker. The solid black line represents the simulated trajectory and the colored rectangular boxes represent the digitized energy deposits (the color code for the energy is represented by the bottom palette). The red line represents the reconstructed trajectory obtained with the hits associated with the primary tracks which are the red dots. With a sample of events that survived these pre-selections, the proton, deuteron, and helium separation was performed.

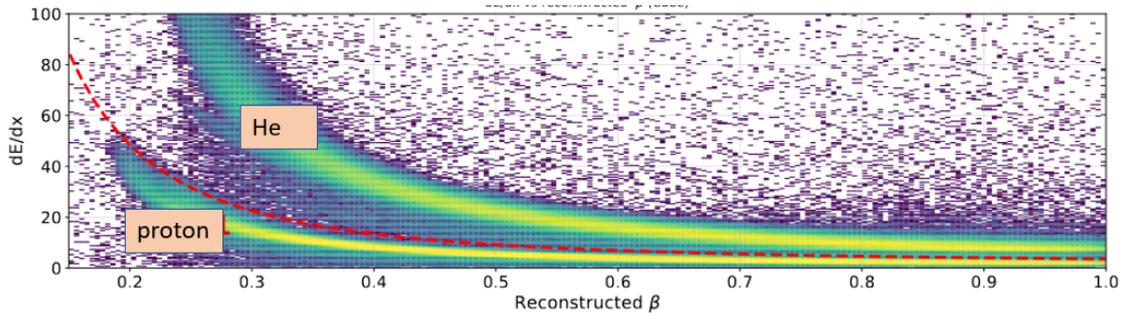

**Figure 2:** The dE/dx deposited in the cube ToF for simulated helium nuclei and protons. The red dotted line represents a selection cut that provides a proton contamination less than 1% over the whole velocity range.

### 3.1 Helium nuclei selection

Helium nuclei selection is performed by means of the specific energy release which, according to the Bethe-Block formula, for particles with electric charge $Z = 2$ is four times higher than $Z = 1$ particles. Figure 2 shows the d$E$/d$X$ measured in the ToF Cube as a function of the reconstructed $\beta$[1]. The dotted red line represents the helium nuclei selection which results in a proton contamination of less than 1%. This cut allows to select helium nuclei with a negligible proton contamination over the whole velocity range.

### 3.2 Deuteron selection

The isotopic separation between proton and deuteron requires additional selections since both are $Z = 1$. Because relativistic hydrogen isotopes release the same d$E$/d$X$ values as a function of $\beta$. The selection of protons and deuterons is performed by exploiting their different kinetic energies and consequently the different ranges inside the tracker.

In order to estimate the kinetic energy and other quantities, a fit of the slow-down profile of the particles is performed. The fit is done using the energy deposits of the primary track. At least three hits are necessary in order to perform the fit, thus only events with at least one hit in the tracker are considered. This allow determining the kinetic energy, the mass of the particle, and the range expressed in cm$^2$/g are obtained. The deuteron selection is performed by combining cuts on these variables. As an example, Figure 3 shows the estimated range as a function of $\beta$ for simulated proton (left panel) and deuteron (right panel). It can be noticed that for the same $\beta$, deuterons are more penetrating and traverse more material. The red line represents the deuteron selection cut. Combining several of these selections, it is possible to obtain a clean deuteron sample.

---

[1]$\beta$ is defined as the ratio between the particle velocity v and the speed of light c, $\beta$=v/c

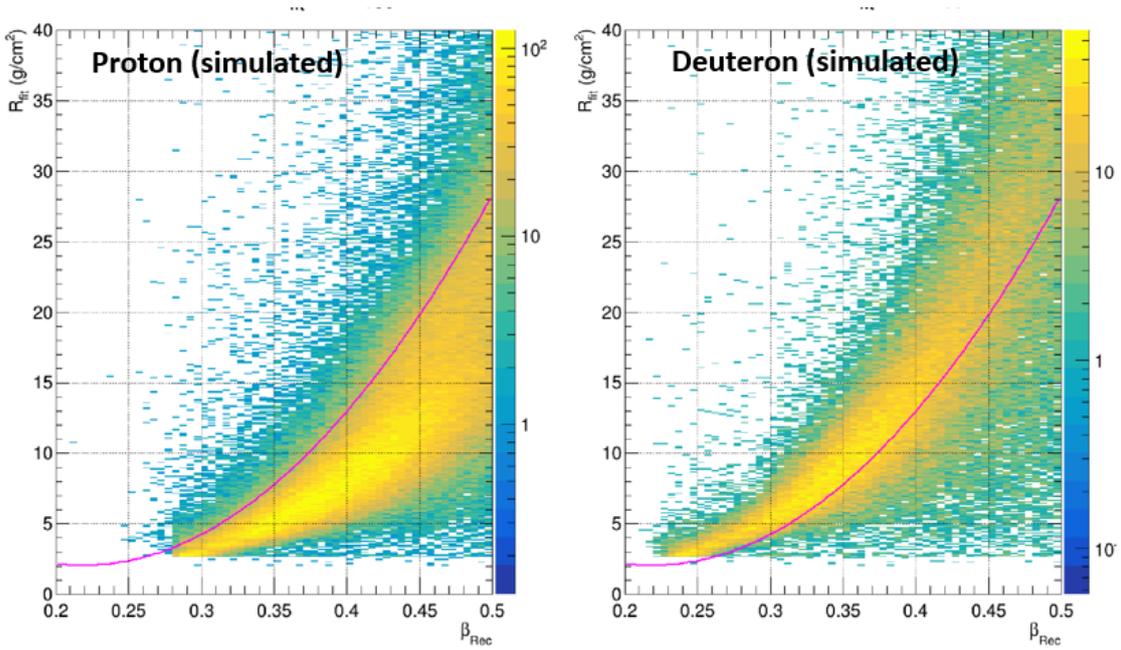

**Figure 3:** Range evaluated from the slow down fit as a function of the reconstructed $\beta$.

Finally, the most relevant quantities for the deuteron identification turned out to be the d$E$/d$X$ release in the Umbrella, the estimated mass of the particle, the estimated kinetic energy and the estimated range. In order to obtain the best performances, all these quantities were used as input of a Recurrent Neural Network (RNN) known as Long-Short Term Memory (LSTM) [10] trained on the samples of simulated protons and deuterons.

## 4. Performance

The performance was studied in terms of acceptance and rejection power. The top panel of Figure 4 shows the deuteron acceptance for the minimum bias trigger configuration (green points) as a function of the simulated kinetic energy per nucleon at the top of the instrument (TOI). The acceptance is almost constant above 30 MeV/$n$ and decreases below this value due to events that stop in the Umbrella ToF. In the same figure, the red points represent the deuteron acceptance after the selection cuts. It has to be noticed that selecting only events crossing the Umbrella and the top Cube decrease the acceptance by more than a factor of 2 since this requirements reduce significantly the angular aperture of the instrument. The blue points represent protons.

The rejection power, defined as ratio of the deuteron over proton acceptance, is represented in the lower panel of Figure 4. The light blue, orange and green arrows in the picture are benchmark values corresponding to a 1%, 3% and 6% of residual proton contamination in the deuteron sample considering typical value of cosmic-ray fluxes near Earth. With this preliminary analysis, it is observed that GAPS allows to measure deuterons with a low proton contamination from ~30 to ~100 MeV/$n$. These energy intervals correspond to ~80 and ~130 MeV/$n$ when expressed in terms of top of the atmosphere (TOA) energy values. The conversion was done running the PLANETOCOSMICS[2] simulation tool and accounting for the atmospheric attenuation for cosmic rays detected at ∼ 37 km.

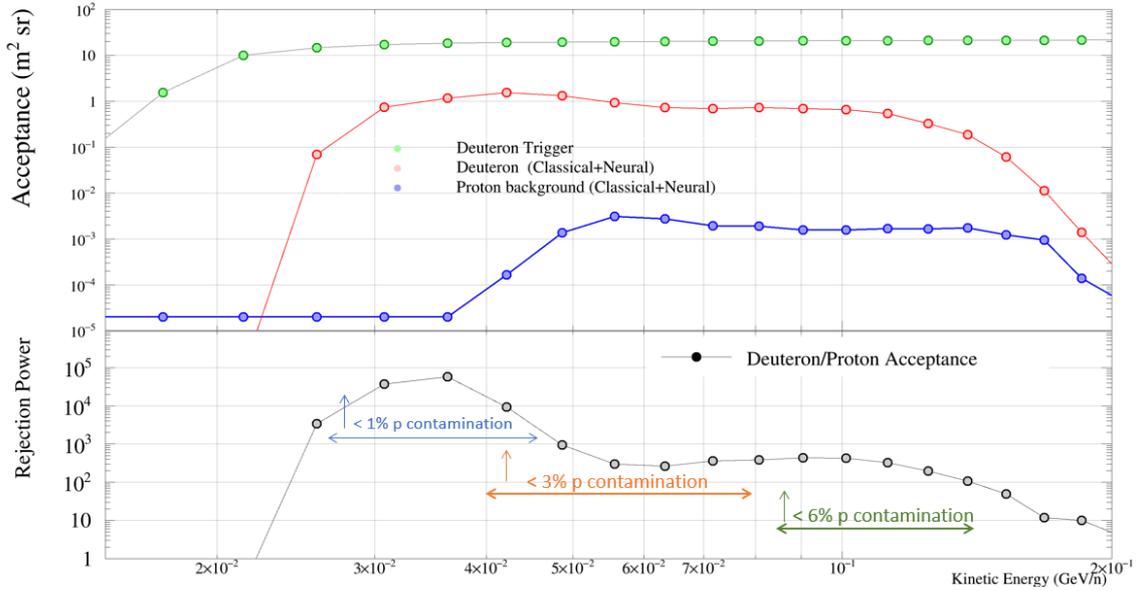

**Figure 4:** Top panel: Deuteron acceptance after the minimum bias trigger (green points) and after the selection cuts and neural network analysis (red points) as a function of the simulated TOI kinetic energy (see text for TOA conversion). The acceptance corresponding to the contaminating protons is represented by the blue dots. Bottom panel: Ratio of the deuteron over proton acceptances (gray points) and proton contamination value according to the cosmic-ray proton and deuteron fluxes near Earth.

Figure 5 shows the accepetance and rejection power for helium nuclei. In this case, the residual proton contamination is well below 1% over the whole kinetic energy range. It has been estimated that according to the $\beta$ resolution, for helium nuclei (and protons) a measurements of the fluxes up to 1 GeV/n (considering bins of 3 times the resolution in $\beta$) is feasable.

## 5. Conclusions and perspectives

This preliminary work demonstrates the capabilities of light nuclei detection and identification with GAPS, an instrument specifically designed for antinucleus detection. With the estimated acceptance and the expected trigger configuration, the fluxes will be measured with a statistical precision of about a percent for deuterons and on the order of per mill for helium nuclei and protons. These measurements will allow to improve the understanding of the cosmic ray light nuclei propagation, especially in the Heliosphere which heavily affects cosmic rays at these energies. Furthermore, models for secondary production in the atmosphere will be tested. The next steps of this work will be:

- Improving the proton rejection power to reach negligible contamination for deuterons at least up to 200 MeV/*n*.

---

[2]http://cosray.unibe.ch/~laurent/planetocosmics/

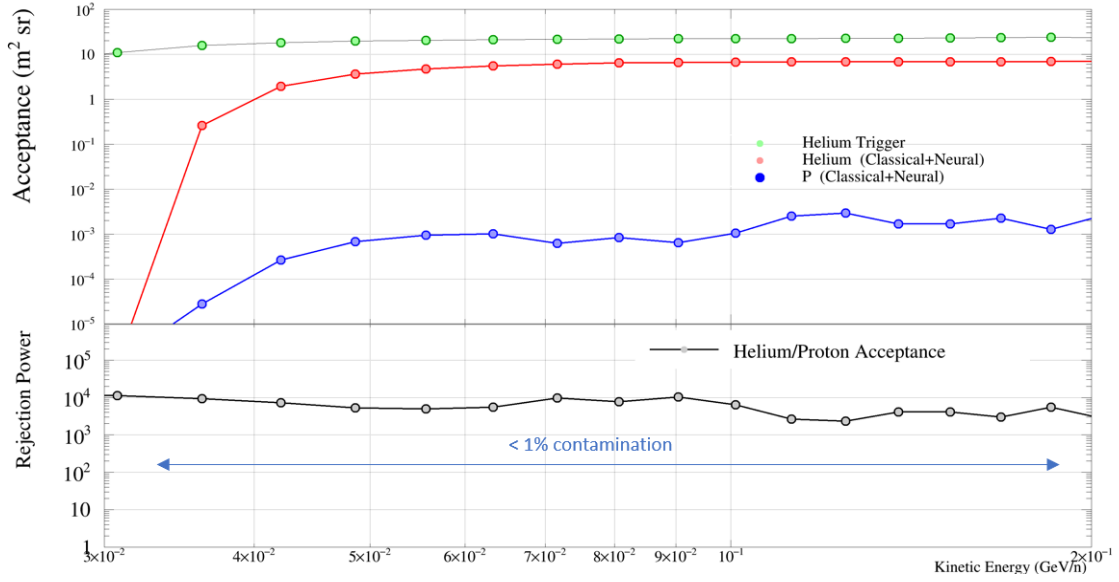

**Figure 5:** Top panel: Helium nuclei acceptance after the minimum bias trigger (green points) and after the selection cuts and neural network analysis (red points) as a function of the simulated TOI kinetic energy. The acceptance corresponding to the contaminating protons is represented by the blue dots. Bottom panel: Ratio of the helium over proton acceptances (gray points) and proton contamination value according to the proton and deuteron flux near Earth.

- Estimate secondary protons, deuterons, and helium nuclei produced in the residual atmosphere. This will help to validate and improve models for secondary production allowing to separate and study the Galactic component.

# References

[1] T. Aramaki et al., Astroparticle Physics 74, 6 2016

[2] K. Mori et al., deexcitation of exotic atoms," ApJ vol. 566, 2002

[3] C.J. Hailey, New J Phys., vol. 11, 2009

[4] T. Aramaki et al., Astropart. Phys., vol. 74, 6–13, 2016

[5] T. Aramaki et al., Astropart. Phys., vol. 59, 12–17, 2014

[6] N. Saffold et al., Astropart. Phys., vol. 130, 102580, 2021

[7] F. Rogers et al., Astropart. Phys., vol. 145, 2023

[8] D.M. Gomez-Coral et al., Phys. Rev. D, 107, 12, 123008, 2023

[9] R. Munini et al., Astroparticle Physics 133, 102640, 2021

[10] S. Hochreiter and J. Schmidhuber, Neural computation. 9. 1735-80, 1997

## Full Authors List: GAPS

T. Aramaki[1], M. Boezio[2,3], S. E. Boggs[4], V. Bonvicini[2], G. Bridges[5], D. Campana[6], W. W. Craig[7], P. von Doetinchem[8], E. Everson[9], L. Fabris[10], S. Feldman[9], H. Fuke[11], F. Gahbauer[5], C. Gerrity[8], L. Ghislotti[15,16], C. J. Hailey[5], T. Hayashi[5], A. Kawachi[12], M. Kozai[13], P. Lazzaroni[15,16], M. Law[5], A. Lenni[3], A. Lowell[7], M. Manghisoni[15,16], N. Marcelli[18], K. Mizukoshi[27], E. Mocchiutti[2,3], B. Mochizuki[7], S. A. I. Mognet[19], K. Munakata[20], R. Munini[2,3], S. Okazaki[27], J. Olson[22], R. A. Ong[9], G. Osteria[6], K. Perez[5], F. Perfetto[6], S. Quinn[9], V. Re[15,16], E. Riceputi[15,16], B. Roach[23], F. Rogers[7], J. L. Ryan[9], N. Saffold[5], V. Scotti[6,24], Y. Shimizu[25], K. Shutt[7], R. Sparvoli[17,18], A. Stoessl[8], A. Tiberio[26,29], E. Vannuccini[26], M. Xiao[23], M. Yamatani[11], K. Yee[23], T. Yoshida[11], G. Zampa[2], J. Zeng[1], and J. Zweerink[9]

[1] Northeastern University, 360 Huntington Avenue, Boston, MA 02115, USA [2] INFN, Sezione di Trieste, Padriciano 99, I-34149 Trieste, Italy [3] IFPU, Via Beirut 2, I-34014 Trieste, Italy [4] University of California, San Diego, 9500 Gilman Dr., La Jolla, CA 90037, USA [5] Columbia University, 550 West 120th St., New York, NY 10027, USA [6] INFN, Sezione di Napoli, Strada Comunale Cinthia, I-80126 Naples, Italy [7] Space Sciences Laboratory, University of California, Berkeley, 7 Gauss Way, Berkeley, CA 94720, USA [8] University of Hawai'i at Mānoa, 2505 Correa Road, Honolulu, Hawaii 96822, USA [9] University of California, Los Angeles, 475 Portola Plaza, Los Angeles, CA 90095, USA [10] Oak Ridge National Laboratory, 1 Bethel Valley Rd., Oak Ridge, TN 37831, USA [11] Institute of Space and Astronautical Science, Japan Aerospace Exploration Agency (ISAS/JAXA), Sagamihara, Kanagawa 252-5210, Japan [12] Tokai University, Hiratsuka, Kanagawa 259-1292, Japan [13] Polar Environment Data Science Center,Joint Support-Center for Data Science Research,Research Organization of Information and Systems,(PEDSC, ROIS-DS), Tachikawa 190-0014, Japan [15] Università di Bergamo, Viale Marconi 5, I-24044 Dalmine (BG), Italy [16] INFN, Sezione di Pavia, Via Agostino Bassi 6, I-27100 Pavia, Italy [17] INFN, Sezione di Roma "Tor Vergata", Piazzale Aldo Moro 2, I-00133 Rome, Italy [18] Università di Roma "Tor Vergata", Via della Ricerca Scientifica, I-00133 Rome, Italy [19] Pennsylvania State University, 201 Old Main, University Park, PA 16802 USA [20] Shinshu University, Matsumoto, Nagano 390-8621, Japan [22] Heliospace Corporation, 2448 6th St., Berkeley, CA 94710, USA [23] Massachusetts Institute of Technology, 77 Massachusetts Ave., Cambridge, MA 02139, USA [24] Università di Napoli "Federico II", Corso Umberto I 40, I-80138 Naples, Italy [25] Kanagawa University, Yokohama, Kanagawa 221-8686, Japan [26] INFN, Sezione di Firenze, via Sansone 1, I-50019 Sesto Fiorentino, Florence, Italy [27] Research and Development Directorate, Japan Aerospace Exploration Agency (JAXA), 2-1-1 Sengen, Tsukuba 305-8505, Japan [28] Research and Development Directorate, Japan Aerospace Exploration Agency, Sagamihara, Kanagawa 252-5210, Japan [29] [29] Universita' degli studi di Firenze, via Sansone 1, I-50019 Sesto Fiorentino, Florence, Italy